\documentclass[11 pt]{article} \textheight=22 true cm
\usepackage{graphicx} 
\usepackage{pslatex}
\usepackage{subfigure}
\usepackage{subcaption}
\usepackage{cite}
\usepackage{graphicx}
\usepackage{amsmath}
\usepackage{amssymb}
\usepackage{color}
\usepackage{breqn}
\usepackage{array}
\usepackage{mathrsfs}
\usepackage{enumerate}
\usepackage{tabularx}   
\usepackage{booktabs}   
\usepackage[colorlinks]{hyperref}
\title{Higher-Dimensional Cosmology in the Framework of Generalized Chaplygin Gas: An Approximation Method}
\author{D. Panigrahi\footnote{ Netaji Nagar Day College, 170/436 N. S. C. Bose Road, Regent Estate, Kolkata  700092, INDIA
\emph{and also} Relativity and Cosmology Research Centre, Jadavpur
University, Kolkata - 70032, India , e-mail:
dibyendupanigrahi@yahoo.co.in, dpanigrahi@nndc.ac.in }, S. Chatterjee \footnote{ New Alipore College (Retd.), Kolkata - 700053, India \emph{and
also} Relativity and Cosmology Research Centre, Jadavpur
University,
Kolkata - 700032, India, e-mail : chat\_sujit1@yahoo.com },  B. C. Paul\footnote{Department of Physics, University of North Bengal, Dist.-Darjeeling, PIN-734013, India, e-mail : bcpaul@associates.iucaa.in}}

\date{}
\begin{document}

\maketitle
\begin{abstract}
In the paper, we have presented a higher-dimensional cosmological model with a generalized Chaplygin-type gas which accommodates the recent cosmic acceleration. We adopt dimensional reduction technique in this model, and obtain new solutions of the higher dimensional dynamical field equations which however reduces to  the Chaplygin gas driven cosmological solutions in $4$  dimensions for  $d =0$.  Using Observed Hubble data from the differential age method, we analyze the higher dimensional cosmological model and estimated constraints of the model parameters. A note worthy aspect of the model is that the resulting field equation is highly nonlinear with respect to the scale factor. However, we investigate  both the dust-dominated and the late accelerating universes in the framework of non-linearity for a given constraints on the model parameters which are relevant to obtain realistic scenario. We obtain cosmological solutions  which are new and interesting which are employed here to probe the late universe.  We have adopted a first-order approximation of the key equation, which permits an exact, time-dependent solution for the scale factors both in the usual 4D and in presence of the extra dimensions. From which we determine the \textit{flip time} of the universe from decelaration to acceleration phase.
The higher dimensional model is found to converge to a $\Lambda$CDM model for a large-scale factor at late time, accommodating the desired feature of an acceleration flip. We also explore the temporal evolution of the deceleration parameter, effective EoS, jerk parameter, etc., in terms of the redshift parameter ($z$). In the higher dimensional cosmological model, the 
   Chaplygin gas parameter $\alpha$  picks up values   less than unity for realistic scenario which is different from pure Chaplygin gas model ($\alpha = 1$) in the usual four dimensional cosmology. The lower value of $\alpha$ is confirmed in accordance with  the observed analysis, indicating  a different  Chaplygin gas parameter  $\alpha \neq 1$ that can accommodate the later accelerating universe.

\end{abstract}

\section{Introduction}\label{sec:intro}
\vspace{0.01cm} \sloppy

Interests have recently revived in gravitational theories that offer explanations of the universe's current accelerated expansion. Since gravitational force is inherently attractive, this observed acceleration challenges our conventional understanding. Nevertheless, detailed analyses of redshifts in Type Ia distant nebulae and cosmic microwave background anisotropy measurements suggest an accelerating expansion.\\

\hspace{-0.6cm} Several explanations have been proposed, including higher-derivative theories~\cite{ua}, a variable cosmological constant in Einstein's field equations~\cite{nas}, axion flavor oscillations~\cite{kalop}, spatial inhomogeneities~\cite{aln, dp0}, scalar fields of a quintessential nature~\cite{pd}, the presence of extra spatial dimensions~\cite{dp1}, and, notably, a Chaplygin-type gas as a matter field~\cite{kp} to accommodate the late acceleration.\\

\hspace{-0.6cm} The authors of this article have recently reported that late-time cosmic acceleration might arise due to a higher-dimensional (HD) phenomenon~\cite{dp, sc1}. Within the framework of higher-dimensional cosmology, we propose that acceleration can be understood as a consequence of additional spatial dimensions, an effect we term a '\emph{dimension-driven}' accelerating model. In this context, the effective Friedmann equations include additional terms due to the presence of extra dimensions, which may be interpreted as a sort of  'fluid' responsible for the observed acceleration. The objective of the approach is to reproduce cosmic acceleration using the geometry of higher-dimensional spacetime itself, without invoking an external scalar field with significant negative pressure.\\

\hspace{-0.6cm}Furthermore, the origin of this extra 'fluid' has a geometric basis, aligning more closely with the principles of general relativity as proposed by Einstein~\cite{einstein} and later expanded among others by Wesson and his collaborators~\cite{wes}. Milton~\cite{milton} demonstrated in an earlier work that while quantum fluctuations in $4D$ spacetime do not account for dark energy, fluctuations in quantum fields, including those involving quantum gravity in extra compactified dimensions, might contribute to dark energy. This notion has led many researchers to explore higher-dimensional space in efforts to unify gravity with other fundamental forces, interpret various brane models, and support the space-time-matter (STM) proposal~\cite{wes}, as well as dimension-driven quintessential models~\cite{pie} in this direction.\\

\hspace{-0.6cm}This investigation is primarily motivated by two considerations. While there are many multidimensional cosmological models in the literature~\cite{bron} and some brane models with generalised Chaplygin-type fluids~\cite{mak, her}, little attention has been given to models that explain cosmic acceleration through extra dimensions themselves  by employing Chaplygin-type matter fields~\cite{ran, sal}.  \\

\hspace{-0.6cm}
This work generalizes our previous study~\cite{dpej}, in which we investigated the pure Chaplygin gas model ($\alpha = 1$) in a higher-dimensional spacetime. It also generalizes the recent work~\cite{dp3}, in which the generalized Chaplygin gas (GCG) model was discussed in four-dimensional spacetime. It  essentially comprises two main parts. We consider a  $(d+4)$-dimensional homogeneous spacetime with two scale factors and a perfect fluid as the source field. Here, a generalized Chaplygin-type matter field is introduced within a higher-dimensional spacetime. However, a closed-form solution of the key Eq.~\eqref{eq:14} is unattainable, as integration yields only an elliptical solution, resulting in a hypergeometric series. Nevertheless, insights can still be drawn from extreme cases, where our analysis shows that the model transits from an initially decelerating phase to an accelerating phase, similar to the 4D case. \\

\hspace{-0.6cm} An intriguing outcome in this context is the emergence of additional terms in the effective equation of state (EoS) at late stages of evolution, originating from the presence of extra dimensions. This result is notably similar to the EoS derived by Guo \emph{et al.}~\cite{guo} for a variable Chaplygin gas model. Depending on the presence of extra dimensions, the universe evolves into either a $\Lambda$CDM or a Phantom-type model. This behavior contrasts sharply with typical $4D$ models, which generally culminate in a de-Sitter phase over time. Although not exactly identical, this finding bears resemblance to ‘\emph{k-essence}’ models, which account for the universe’s current acceleration across a broad range of initial conditions without requiring fine-tuning or anthropic reasoning. \\

\hspace{-0.6cm}Here we apply a $\chi^2$ minimization technique to obtain constraints from cosmological observations. Using Type Ia Supernova data and predictions from CMB and BAO, we constrain our cosmological models. By defining a total $\chi^2$ function, we analyze the models with the $(H(z) - z)$ OHD data. We are constraining the parameters $\Omega_m$, $\alpha$ and $m$ for $d = 0, 1 ~\& ~ 2$ using Hubble-$57$ data. Here we find $m>0$ which indicates  possibility of dimensional reduction mechanism. Again, the value of $\alpha$ is much less than unity, which disagrees with the pure Chaplygin gas model ($\alpha = 1$) in higher-dimensional spacetime. \\

\hspace{-0.6cm}It is worth noting that the desirable feature of dimensional reduction of extra space is feasible in this model. However, we cannot fully account for the impact of compactifying extra dimensions on the present acceleration or the universe’s scale factor evolution, as the key Eq.~\eqref{eq:14} does not yield an explicit solution. Thus, we restrict our analysis to extremal values. This limitation may be addressed through an \emph{approximate method}~\cite{dp2}, where higher-order terms in the binomial expansion on the RHS of Eq.~\eqref{eq:14} are neglected. The rationale behind this approximation is that in the zero-pressure era, the $4D$ scale factor should be sufficiently large, making it reasonable to consider only the first-order terms of the binomial expansion, as shown in Eq.~\eqref{eq:33}.
We obtain here an exact solution, allowing us to study an explicit time-dependent solution.

\section {Higher Dimensional Field Equations :}\label{sec:fe}
We consider the Einstein-Hilbert action for the $(d + 4 )$-dimensional spacetime as
\begin{equation}\label{eq:1}
\mathscr{A} = \frac{1}{16 \pi G_{d+4}}\int \textbf{d}^{d+4}x \sqrt{|g|}R
\end{equation}
where $g_{ij}$ is the metric tensor in $(d+4)$-dimensional spacetime where $i, j$ are $(0,1,2,3, .....d)$, $R_{ij}$ and $R$ are the Ricci tensor and Ricci scalar respectively. Varying the above action we get the Einstein field equation in $(d+4)$-dimension is given by
\begin{equation}\label{eq:2}
R_{ij} - \frac{1}{2} g_{ij} R = \kappa T_{ij}
\end{equation}
We  consider the line element of a (d+4)-dimensional homogeneous  spacetime as
\begin{eqnarray}\label{eq:3}
  ds^{2} &=&
  dt^{2}-a^{2}(t)\left(\frac{dr^{2}}{1-r^{2}}+r^{2}d\theta^{2}+
  r^{2}sin^{2}\theta d\phi^{2}\right) - b^{2} (t)\gamma_{\mu \nu}dy^{\mu}dy^{\nu}
\end{eqnarray}
here $y^{\mu}$ ($\mu, \nu = 4,....        , 3+d$) are the extra
dimensional spatial coordinates and the scale factors in 3D and extra dimensions
are $a(t) $ and $b(t) $  which  depend on time only  and the compact manifold is described by the metric
$\gamma_{\mu \nu}$. We consider the manifold $M^{1}\times S^{3}\times S^{d}$
the symmetry group of the spatial section is $O(4) \times O(d+1)
$. The stress tensor whose form will be determined by Einstein's
equations must have the same invariance leading to the energy
momentum tensor as \cite{rd}
\begin{equation}\label{eq:4}
T_{00}=\rho~,~~T_{ij}= -p(t)g_{ij}~,~~ T_{\mu \nu}=-p_{d}(t)g_{\mu \nu}
\end{equation}
where the rest of the components vanish. Here $p$ is the isotropic
3-pressure and $p_{d}$, that in the extra dimensions.
The field equations are as follows

\begin{eqnarray}
\rho &=& 3 \frac{\dot{a}^2}{a^2}+\frac{1}{2}d(d-1) \frac{\dot{b}^2}{b^2} + 3 d \frac{\dot{a}}{a}\frac{\dot{b}}{b}  ~~~~~\label{eq:4a}\\
 - p &=& 2\frac{\ddot{a}}{a} + \frac{\dot{a}^2}{a^2} + d \frac{\ddot{b}}{b} + \frac{1}{2} d(d-1)\frac{\dot{b}^2}{b^2} + 2 d \frac{\dot{a}}{a}\frac{\dot{b}}{b} ~~~~~\label{eq:4b}\\
- p_{d} &=&  3\frac{\ddot{a}}{a} + 3\frac{\dot{a}^2}{a^2} + (d -1) \frac{\ddot{b}}{b} + \frac{1}{2} (d-1)(d-2) \frac{\dot{b}^2}{b^2} + 3 d(d-1)\frac{\dot{a}}{a}\frac{\dot{b}}{b} ~~~~~\label{eq:4c}
\end{eqnarray}
To make it convenient for dimensional reduction, let us consider
\vspace{0.5cm}

\begin{equation}\label{eq:5}
b(t) = a(t)^{-m}
\end{equation}
here $m$ is any positive constant.  For the matter
field we here assume an equation of state given by the Generalised Chaplygin type of gas in $3D$ space only ~\cite{kp} which is
\begin{equation}\label{eq:6}
p= - \frac{B}{ \rho^{\alpha}}
\end{equation}
In the generalized Chaplygin gas (GCG) framework, the parameter $\alpha$ is an arbitrary constant that plays a crucial dynamical role and significantly extends the physical scope of the original Chaplygin gas model. The introduction of $\alpha$ improves both the physical consistency and the observational viability of the model by controlling the smoothness and timing of the transition from the matter-dominated era  to the dark-energy-dominated phase, thereby allowing a more realistic description of the late-time accelerated expansion of the Universe. Unlike the original Chaplygin gas model, which corresponds to the fixed choice $\alpha=1$, the generalized model with small values of $\alpha$ provides a substantially better fit to observational data from SNe~Ia, BAO, $H(z)$ measurements, CMB, and OHD~\cite{lu}. At the perturbative level, the square of the adiabatic sound speed, $c_s^2 = \alpha\frac{ B}{\rho^{\alpha+1}}$, must remain subluminal ($c_s^2 < c^2$), leading to the constraint $0 < \alpha < c^2 \frac{\rho^{\alpha+1}}{B}$. As the Universe evolves and the energy density $\rho$ decreases, the upper bound on $\alpha$ becomes progressively tighter, naturally favoring smaller values of $\alpha$ at late times, in agreement with our previous findings~\cite{dp3}. Because the sound speed depends explicitly on the Chaplygin gas parameter $\alpha$, smaller values of $\alpha$ naturally suppress excessive late-time sound speeds, preventing the development of unphysical oscillations and instabilities in the matter power spectrum that are characteristic of the original Chaplygin gas model. Furthermore, the parameter $\alpha$ governs how pressure responds to the energy density and therefore directly influences key cosmological quantities such as the Hubble expansion rate, the deceleration parameter, and the redshift of transition from deceleration to acceleration.
Finally, allowing $\alpha$ to vary provides the necessary flexibility to achieve simultaneous consistency with multiple observational probes, while in the limit $\alpha \rightarrow 0$ the GCG model smoothly approaches a $\Lambda$CDM behavior, ensuring compatibility with current observational constraints.

Now, using Eqs.~\eqref{eq:4a}, \eqref{eq:4b}, \eqref{eq:4c}, and \eqref{eq:5}, we obtain the following set of equations~\cite{dp} expressed in terms of the Hubble parameter $H=\dot a/a$:
\begin{eqnarray}
\rho & =& \frac{k}{2} H^2\label{eq:7} ~~~~~~~~~~~~~~~~~~~~~~~~~~~~~~~~~~~~~~~~~~~~~~~~~~~\\
 -p & = & (2-dm)(\dot{H} +H^2) + \frac{1}{2}[m^{2}d( d+1) +
2(1-dm)]H^2\label{eq:8} \\
-p_{d} &= &(3-dm + m)(\dot{H} +H^2)+\frac{1}{2}[m(d-1)(dm-4)+6]H^2 \label{eq:9}
\end{eqnarray}
where $k = m^{2} d(d-1) + 6(1-dm)$. For a positive energy density, $k$ must be greater
than zero which implies $m < \frac{3d-\sqrt{3d(d+2)}}{d(d-1)}$
 or, $m > \frac{3d+\sqrt{3d(d+2)}}{d(d-1)}$.
\vspace{0.5 cm}
The conservation equation is given by
\begin{equation}\label{eq:10}
\dot{\rho} + \{3(\rho + p) -
dm (\rho + p_{d})\} H = 0
\end{equation}

Now using Eqs.~\eqref{eq:8}, \eqref{eq:9} \& \eqref{eq:10} we get
\begin{eqnarray}\label{eq:11}
\dot{\rho} + \frac{k}{(2-dm)}H\left[\left\{1
+ \frac{2 dm (m+1)}{k}\right\}\rho - B\rho^{-\alpha}\right] =0
\end{eqnarray}
Solving Eq.~\eqref{eq:11} we finally get
\begin{eqnarray}\label{eq:12}
\rho = \left[ \frac{Bk}{M} +
c (1+z)^{\frac{2 M(1+\alpha)}{(2-dm)}}\right]^{\frac{1}{1+ \alpha}}
\end{eqnarray}
where
\begin{equation}\label{eq:13}
M = k + 2 d  m (m+1)
\end{equation}
 and
$c$ is the integration constant.  Based on physical considerations, the restriction on $m$ as $m <\frac{3d-\sqrt{3d(d+2)}}{d(d-1)}$ for $ d \neq 1$ otherwise $m <1$ for $ d = 1$. In addition, the condition $2 >dm$ must also be satisfied. On the other hand, $m$ is not relevant for $d = 0$
since it represents the index of extra spatial dimensions. It is worth mentioning that a detailed analysis of this was provided in our previous work~\cite{dpej} within the context of pure Chaplygin gas cosmology. The last term on the right-hand side of Eq.~\eqref{eq:13} arises from higher-dimensional contributions, which are absent in the $4D$ case ($d = 0$). Consequently, the density of the universe at the present epoch is lower in the framework of a higher-dimensional universe compared to a 4-dimensional universe. The following points are noteworthy:

(\emph{i}) For $m= -1$, the universe evolves with
$b(t)=a(t)$, leading to expansion across all dimensions, making the desirable dimensional reduction of extra space unattainable.

(\emph{ii}) For $m=0$, the universe exhibits flat extra space in $(d+3)$ dimensions, resembling the scenario in a
$4D$ universe, as noted in  Ref. ~\cite{ud}. In fact, this similarity directly follows from Campbell's theorem, which states that any analytic $N$-dimensional Riemannian manifold can be locally embedded into a higher-dimensional Ricci-flat manifold
 ~\cite{tavako}.

(\emph{iii})  For  $d =0$, the solutions correspond to the well-known features of $4D$ cosmology~\cite{dp3}. \vspace{0.5cm}

Now using Eqs.~\eqref{eq:7}  \&  \eqref{eq:12}, we get
\begin{eqnarray}\label{eq:14}
H^2 = \frac{\dot{a}^2}{a^2}=\frac{2}{k} \left[ \frac{Bk}{M} + c (1+z)^{\frac{2 M(1+\alpha)}{(2-dm)}}\right]^{\frac{1}{1+\alpha}}
\end{eqnarray}

The solution of the Eq.~\eqref{eq:14} in terms of scale factor  $a(t)$ cannot be obtained in a closed form  because integration yields an elliptical solution which gives  hypergeometric series. However under extremal condition  significant information may be obtained from  the Eq.~\eqref{eq:14} as briefly discussed in the next section.

\section { Cosmological dynamics : }\label{sec:cd}
Now we would like to  discuss the cosmological behaviour of the the generalised Chaplygin gas equation of state in higher dimensional spacetime.
We have expressed the relevant equations as a function of redshift ($z$). We have considered the value of parameters to draw the graphs from the analysis of observational data and discuss the cosmological dynamics.
\vspace{0.5 cm}
\subsection{Deceleration Parameter:}\label{sec:q}

At the early stage of the cosmological evolution
when the red shift $z$ of the universe was high enough, the second term of the right hand side of the Eq.~\eqref{eq:14} dominates which has been discussed in the literature ~\cite{bento} for the case of $4D$ universe.
Now the deceleration parameter $q$ for generalised Chaplygin gas in higher dimensional spacetime is given by,

\begin{equation}\label{eq:15}
q = \frac{d}{dt} \left(H^{-1}
\right) - 1 =  -1 + \frac{M}{2(2-dm)} -  \frac{Bk}{2(2-dm)} \frac{1}{\rho^{1+\alpha}}
\end{equation}
where $H$ is the Hubble parameter. To constrain the parameters, let us consider $\frac{Bk}{M} = \frac{c (1 - \Omega_m)}{\Omega_m}$. Now with the help of the Eqs.~\eqref{eq:12} and \eqref{eq:15}, we get

\begin{equation}\label{eq:16}
q = -1 + \frac{M}{2(2-dm)} -\frac{M}{2(2-dm)}\frac{1-\Omega_m}{\Omega_m}\left[\frac{1-\Omega_m}{\Omega_m} + (1+z)^{\frac{M(1+\alpha)}{2-dm}}\right]^{-1}
\end{equation}
Again at flip time, {\it i.e.} when $q = 0$ the redshift parameter  $z_{f}$ becomes
\begin{equation}\label{eq:17}
z_{f} = \left[\frac{1-\Omega_m}{\Omega_m} \frac{2(2-dm)}{M-2(2-dm)}\right]^{\frac{2-dm}{M(1+\alpha)}} - 1
\end{equation}
Here $z_f$ denotes the redshift at which the deceleration parameter changes sign.
For the universe to be accelerating at the present epoch (\textit{i.e.}, at $z=0$), one requires $z_f>0$, which leads to the condition
\begin{equation}
\Omega_m < \frac{2(2-dm)}{M} < 1 .
\end{equation}
In the special case $d=0$, this condition reduces to $\Omega_m < 2/3$, which is fully consistent with current observational constraints indicating $\Omega_m \lesssim 0.3$. From our observational analysis, we obtain $\Omega_m = 0.2443$, which satisfies this bound and supports an accelerating universe at the present epoch, in agreement with observations.

For $0<m<1$, the transition redshift $z_f$ decreases monotonically with increasing dimension $d$, implying that higher-dimensional contributions delay the flip from decelerated to accelerated expansion. In addition, the generalized Chaplygin gas parameter $\alpha$ plays a decisive role in determining the epoch of this flip. While the original Chaplygin gas model ($\alpha=1$) predicts a comparatively late transition, smaller values of $\alpha$ significantly enhance $z_f$ across all dimensions, leading to an earlier and observationally favored flip to cosmic acceleration.

Although cosmic acceleration is a recent phenomenon, this behavior does not conflict with the requirement of small $\alpha$ at late times imposed by the causality condition $c_s^2<c^2$. A smaller $\alpha$ enhances the transition redshift $z_f$ only within the late-time regime, shifting the flip from deceleration to acceleration from an unrealistically late epoch (for example, $z_f \lesssim 0.4$ for $\alpha=1$ in the four-dimensional case) to a more observationally consistent range ($z_f \sim 0.6$--$1$). In this context, it is worth noting that Dubey \textit{et al.}~\cite{du} found $z_f = 0.62$, while Pacif \textit{et al.}~\cite{pa} obtained $z_f = 0.7$ in four-dimensional spacetime. Thus, smaller values of $\alpha$ ensure physical stability, subluminal sound speed, and a realistic description of late-time cosmic acceleration, without shifting the acceleration phase to the early Universe.

\begin{table}[h!]
\centering
\caption{\small Transition redshift $z_f$ for different dimensions and values of the generalized Chaplygin gas parameter $\alpha$.}
\label{t_zf}
{\fontsize{08}{10}\selectfont
{
\begin{tabular}{ccccc}
\hline
Dimension ($d$ ) & $\alpha$ & $m$ & $z_f$ & \\
\hline
0 & 0.030  & --     & 0.804 & \cite{dp3} \\
1 & 0.089 & 0.268  & 0.465 &  \\
2 & 0.126 & 0.183  & 0.372 &  \\
0 & 1.000   & --     & 0.355 & \cite{dp3} \\
1 & 1.000   & 0.540     & 0.350 & \cite{dpej}\\
\hline
\end{tabular}
}
}
\end{table}

From Table-\ref{t_zf}, it is evident that the transition redshift $z_f$ is sensitive to both the spatial dimension $d$ and the parameter $\alpha$. In the standard four-dimensional case ($d=0$), smaller values of $\alpha$ lead to a significantly larger $z_f$ compared to the original Chaplygin gas model, indicating an earlier—yet still late-time—flip to accelerated expansion. As the dimension $d$ increases, $z_f$ decreases systematically, confirming that higher-dimensional effects tend to delay the acceleration phase for $0<m<1$. Overall, these results demonstrate that the generalized Chaplygin gas model with small $\alpha$ provides a physically viable and observationally consistent framework for explaining the recent accelerated expansion of the Universe.

As the universe expands, the energy density $\rho$ decreases with time. Consequently, the last term in Eq.~\eqref{eq:15} grows in magnitude, leading to a change in the sign of the deceleration parameter when the energy density reaches a critical value. This critical density corresponds to the transition redshift $z_f$ discussed above, marking the flip from decelerated to accelerated expansion, and is given by
\begin{equation}\label{eq:18}
\rho = \rho_f = \left[ \frac{Bk}{M - 2(2-dm)} \right]^{\frac{1}{1+\alpha}} .
\end{equation}
It is evident that for $M > 2(2-dm)$, the Universe remains dominated by normal matter prior to the transition. This condition is therefore consistent with the existence of a realistic and physically acceptable transition redshift $z_f$.

\begin{figure}[ht]
\begin{center}
  \includegraphics[width=8 cm]{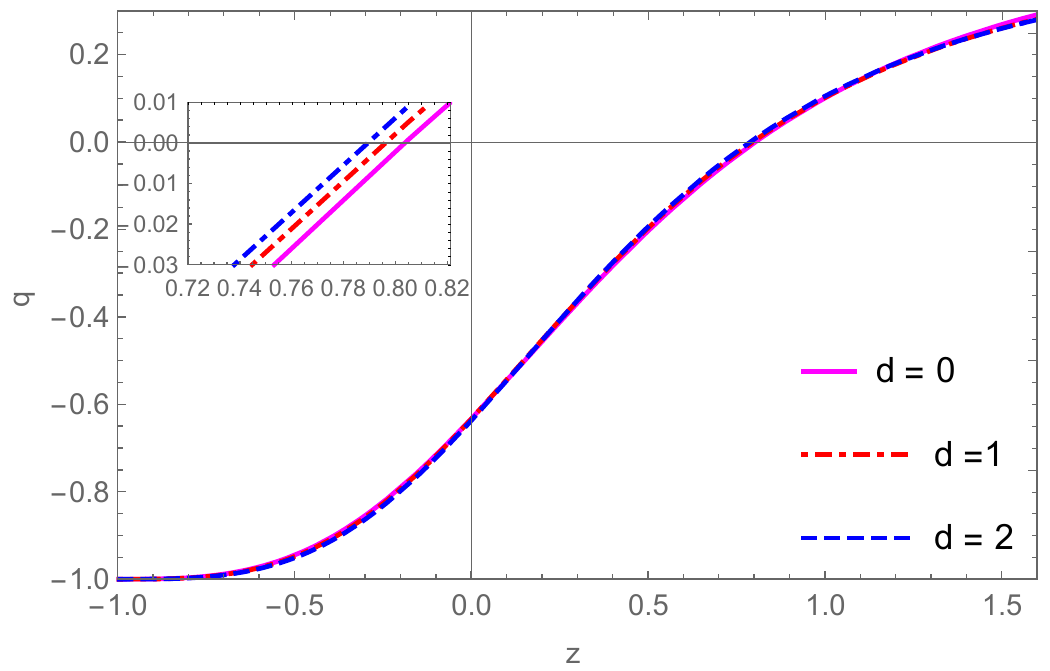}
  \caption{
  \small\emph{$q$ vs $z$ for $H_0 = 71$ using Eq.~\eqref{eq:16}}\label{fig:qzh71}
  }
\end{center}
\end{figure}

Now we discuss the extremal cases in order to further understand the evolution of the Universe.\\
\vspace{0.5 cm}
\textbf{CASE~A :}In the early phase, when the redshift $z$ is very large, Eq.~\eqref{eq:16} reduces to
\begin{equation}\label{eq:19}
q = -1 + \frac{M}{2(2-dm)}
= \frac{1 - dm}{2} + \frac{dm(m+1)}{2(2-dm)} .
\end{equation}
This expression corresponds to a dust-dominated universe for the generalized Chaplygin gas.
For $d=0$, one obtains $q=\frac{1}{2}$, which is the standard result for a four-dimensional
matter-dominated Universe and is in excellent agreement with the well-known $4D$ cosmology.
The positive value of $q$ confirms that the Universe is in a decelerating phase during this epoch.
For $d=1$, Eq.~\eqref{eq:19} yields $q \simeq 0.47$, while for $d=2$ one finds $q \simeq 0.45$.
Both values remain positive and are close to the standard matter-dominated value $q=\frac{1}{2}$,
indicating that the early universe continues to be in a decelerated, dust-dominated phase even in
the presence of extra dimensions. Although the magnitude of $q$ is slightly reduced compared to
the four-dimensional case, the qualitative behavior of early-time cosmological evolution is preserved.\\
\vspace{0.5 cm}
\textbf{CASE~B :}
At the present epoch, when the redshift $z=0$, Eq.~\eqref{eq:16} reduces to
\begin{equation}\label{eq:20}
q = -1 + \frac{M \Omega_m}{2(2-dm)} .
\end{equation}
Since $0<\dfrac{M \Omega_m}{2(2-dm)}<1$, the deceleration parameter $q$ is negative, which represents an accelerating universe at the present phase, as also illustrated in Fig.-\ref{fig:qzh71}.

For the standard four-dimensional spacetime ($d=0$), we obtain $q \simeq -0.625$, which lies between $-1$ and $0$ and thus confirms the present accelerated expansion.
For $d=1$, the corresponding value is $q \simeq -0.64$, while for $d=2$ one finds $q \simeq -0.65$.
All these values remain negative and close to each other, indicating that the universe undergoes accelerated expansion at the present epoch even in the presence of extra dimensions.
The small variation of $q$ with increasing $d$ shows that higher-dimensional effects slightly modify the strength of acceleration but do not change its qualitative nature.
\\
\textbf{CASE~C :}  In the later epoch of evolution, \textit{i.e.}, when the universe has reached a sufficiently large size, we obtain from Eq.~\eqref{eq:16}
\begin{equation}\label{eq:21}
q=-1 ,
\end{equation}
which represents an exact $\Lambda$CDM behavior and is dimension independent.
One possible reason for this behavior is that at this stage the contribution of matter becomes negligible and the dynamics is completely dominated by the dark-energy–like component, so that the effects of extra spatial dimensions are no longer observable.

Fig.-\ref{fig:qzh71} clearly shows that the flip from decelerated to accelerated expansion occurs at a later redshift for higher dimensions. This implies that the presence of extra spatial dimensions tends to retard the onset of cosmic acceleration.
In addition, the generalized Chaplygin gas parameter $\alpha$ also plays an important role in controlling the timing of this flip: smaller values of $\alpha$ shift the transition to higher redshifts, leading to an earlier and smoother approach to the accelerated phase, whereas the original Chaplygin gas case ($\alpha=1$) produces a comparatively delayed transition.
Thus, both the dimensionality of spacetime and the value of $\alpha$ influence the detailed evolution of cosmic acceleration, while the late-time attractor remains the exact $\Lambda$CDM limit with $q=-1$.

\subsection{Effective Equation of State:}\label{sec:eos}

\begin{figure}[ht]
\begin{center}
  \includegraphics[width=8cm]{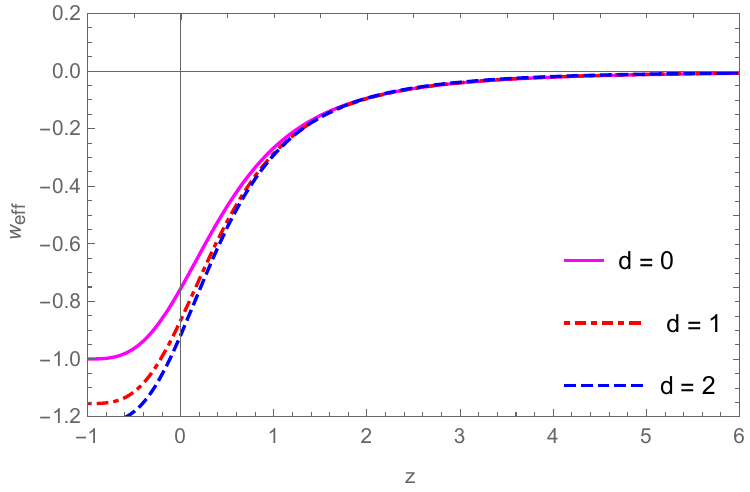}
  \caption{\small\emph{Evolution of the effective equation-of-state parameter $w_{\text{eff}}$ with redshift $z$ obtained from Eq.~\eqref{eq:22}.}}
  \label{fig:wzh71}
\end{center}
\end{figure}

Now the \emph{effective} equation of state (EoS) parameter using Eqs.~\eqref{eq:6} and \eqref{eq:12} as
\begin{equation}\label{eq:22}
w_{\text{eff}} = \frac{p}{\rho}
= - \frac{M(1-\Omega_m)}
{k \left[1 - \Omega_m + \Omega_m (1+z)^{\frac{M(1+\alpha)}{2-dm}}\right]} .
\end{equation}
To understand the cosmological implications of Eq.~\eqref{eq:22}, we now examine the behavior of the effective equation-of-state parameter in different evolutionary phases of the Universe.\\
\textbf{Case A: Early Universe}:
At very high redshift ($z \gg 1$), corresponding to the early phase of the universe, the second term in the denominator of Eq.~\eqref{eq:22} dominates, and the effective EoS reduces to $ w_{\text{eff}} \rightarrow 0 $, which characterizes a dust-dominated Universe. This behavior is also illustrated in Fig.~\ref{fig:wzh71}.\\
\textbf{Case B: Present Epoch}:
At the present epoch ($z=0$), Eq.~\eqref{eq:22} reduces to
\begin{equation}\label{eq:23}
w_{\text{eff}}
= -\left[1 + \frac{2dm(m+1)}{k}\right](1-\Omega_m) ,
\end{equation}
indicating an accelerating universe at present. For the four-dimensional case ($d=0$), this yields $w_{\text{eff}} \approx -0.75$, which is consistent with current observational constraints.\\
\textbf{Case C: Late-Time Universe}:
In the late stage of cosmic evolution, corresponding to a sufficiently large scale factor (or equivalently $z \rightarrow -1$), the effective EoS obtained from Eq.~\eqref{eq:22} approaches
\begin{equation}\label{eq:24}
w_{\text{eff}} = -1 - \frac{2dm(m+1)}{k} .
\end{equation}
For $d=0$, Eq.~\eqref{eq:24} reduces to $w_{\text{eff}}=-1$, reproducing the exact behavior of the $\Lambda$CDM model. The evolution of the effective equation-of-state parameter $w_{\text{eff}}$ is primarily governed by the generalized Chaplygin gas parameter $\alpha$ through the redshift-dependent exponent in Eq.~\eqref{eq:22}. In the original Chaplygin gas limit ($\alpha=1$), the larger exponent keeps $w_{\text{eff}}$ close to zero over a broad redshift interval, thereby extending the matter-dominated era and yielding a comparatively late onset of accelerated expansion. In contrast, for $\alpha<1$, the reduced exponent enables $w_{\text{eff}}$ to become negative at higher redshifts, leading to an earlier and smoother transition to acceleration, in better agreement with observational bounds.

At late times, the effective EoS approaches a constant value, and the cosmic fluid behaves as $p\simeq -\rho$, corresponding to a universe dominated by a cosmological constant. Accordingly, the deceleration parameter approaches $q=-1$, confirming accelerated expansion. While the presence of extra dimensions ($d\neq0$) introduces additional terms that may drive $w_{\text{eff}}<-1$, indicating a phantom-like regime, the four-dimensional limit ($d=0$) naturally recovers the standard $\Lambda$CDM behavior. This suggests that extra-dimensional effects become dynamically negligible in the late-time evolution of the universe.

Fig.-\ref{fig:wzh71} shows that $w_{\text{eff}}$ evolves from a dust-like phase at high redshift to a dark energy-dominated phase at late times, with $w_{\text{eff}}<0$ at the present epoch. The generalized Chaplygin gas model thus provides a unified and observationally viable description of the transition from decelerated to accelerated expansion, with smaller values of $\alpha$ favored both by data and by the causality requirement $c_s^2<c^2$.

\vspace{0.5 cm}
\subsection{Jerk parameter:}\label{sec:jerk1}
\vspace{0.3 cm}

The jerk parameter, a dimensionless third derivative of the scale factor $a(t)$ with respect to cosmic time $t$,  can provide us the simplest approach to search for departures from the concordance $\Lambda$CDM model. It is defined as ~\cite{blan,rap}

\begin{equation}\label{eq:25}
j = \frac{dq}{dt} = - \frac{1}{a H^3}\frac{d^3a}{dt^3}
\end{equation}

Now, the jerk parameter $j$ can be written  in terms of deceleration parameter $q $ as

\begin{equation}\label{eq:26}
 j(q) =   q(2q + 1) + (1 + z)\frac{dq}{dz}
\end{equation}

Blandford \textit{et al}. ~\cite{blan} described how the jerk parameterization provides an alternative and a convenient method to describe cosmological models close to concordance $\Lambda$CDM model. A powerful
 feature of $j$ is that for the $\Lambda$CDM model $j = 1 $ (constant) always. It should be noted here that
 Sahni \textit{et al.} ~\cite{sah, alam} drew attention to the importance of $j$ for discriminating different dark
  energy models, because any deviation from the value of $j = 1$ (just as deviations from the effective  equation of state parameter $w_{\text{eff}} =-1$ do in more standard dynamical approaches)    would favour a non-$\Lambda$CDM model. The simplicity of the jerk formalism thus enables us
    to constrain the departure from the $\Lambda$CDM value in an effective manner.
Now using Eqs.~\eqref{eq:16} and \eqref{eq:26} we get

\begin{dmath}\label{eq:27}
 j(q) =   \left \{ \frac{M}{2(2-dm)} - \frac{ \frac{M}{2(2-dm)} \frac{1 - \Omega_m}{\Omega_m}}{\frac{1 - \Omega_m}{\Omega_m} +(1+z)^{\frac{M(1+\alpha)}{2-dm}}} - 1\right\} \times \\
 \left[2\left\{\frac{M}{2(2-dm)}-  \frac{ \frac{M}{2(2-dm)} \frac{1 - \Omega_m}{\Omega_m}}{\frac{1 - \Omega_m}{\Omega_m} +(1+z)^{\frac{M(1+\alpha)}{2-dm}}} \right\} - 1 \right] \\
 +  \frac{ \frac{M}{2(2-dm)} \frac{1 - \Omega_m}{\Omega_m}}{\left\{\frac{1 - \Omega_m}{\Omega_m} +(1+z)^{\frac{M(1+\alpha)}{2-dm}}\right\}^{2}}
  \frac{M(1+\alpha)}{2 -d m}(1+z) ^{\frac{M(1+\alpha}{2-dm}}
\end{dmath}
The Eq.~\eqref{eq:27} is so involved, we can not analytically obtain any useful conclusion, so we have discussed this phenomena in graphical approach.

\begin{figure}[ht]
\begin{center}
  \includegraphics[width=8 cm]{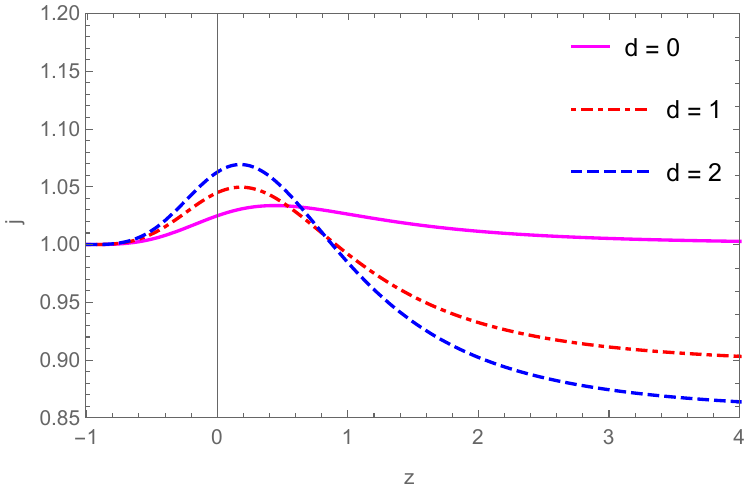}
  \caption{
  \small\emph{$ j$ vs $z$ using equation \eqref{eq:27}  }\label{fig:jzh71}
    }
\end{center}
\end{figure}

The Fig.-\ref{fig:jzh71} shows three different lines, each corresponding to dimensions $d = 0$, $d = 1$ and $d = 2$ for  $H_0 = 71~km s^{-1}Mpc^{-1}$~\cite{sha}. Later they  converge at $j = 1$ in each case, which corresponds to $\Lambda$CDM at future cosmic time. These are in good agreement  with the present observational analysis of our universe. It once again suggest that the cosmological evolution favours the insignificance influence of extra dimension in the late world.

\vspace{0.5 cm}
\subsection{Observational Constraints on  the  Model Parameters:}\label{sec:obs}
\vspace{0.3 cm}

In this section the  Hubble-$57$ data ~\cite{sha} will be used to  analyze cosmological model estimating the constraints imposed on the model parameters. Here, the value of Hubble parameter $H(z)$ at certain redshift $z$ can be measured with two methods
(i) $H(z)$ estimations from differential ages (DA) $\,triangle t$ of galaxies. (ii) extraction $H(z)$ from line-of-sight BAO data  including analysis of correlation functions of luminous red galaxies. The detailed data table was presented in our recent article~\cite{dp3}.

\begin{figure}[ht]
\begin{center}
  \includegraphics[width=8 cm]{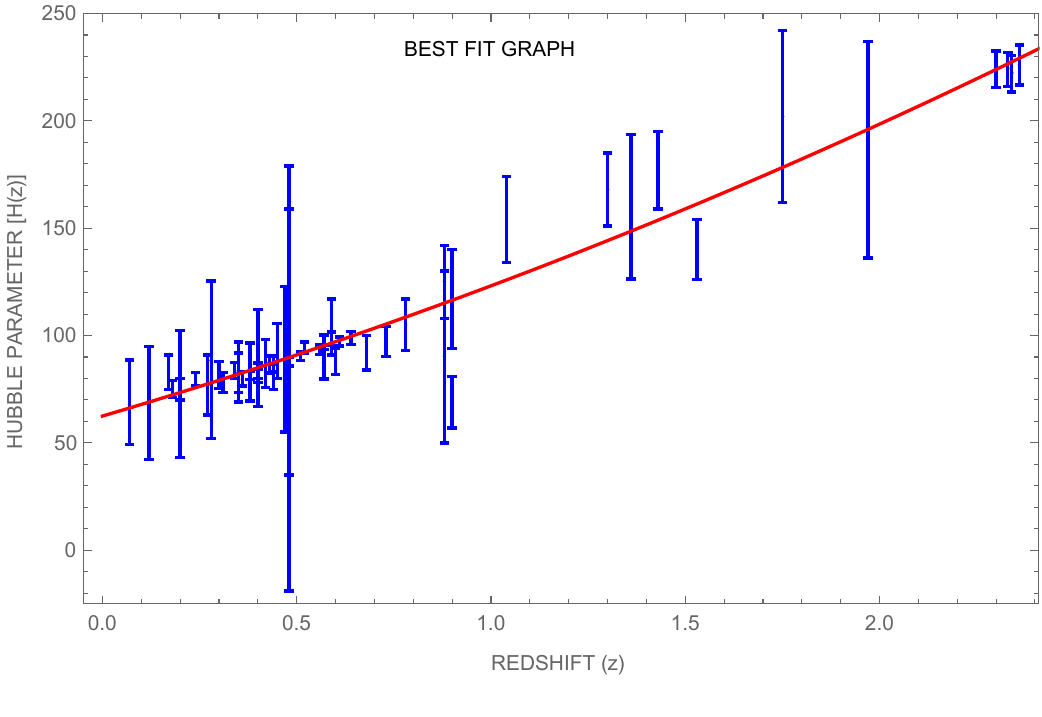}
  \caption{
  \small\emph{The variation of Hubble parameter $H(z)$ with Redshift  $z$  }\label{fig:BFG}
    }
\end{center}
\end{figure}

The Hubble parameter depending on the differential ages as a function of  redshift $ z$ can be written in the form of

\begin{equation}\label{eq:28}
H(z) = -\frac{1}{1+z} \frac{dz}{dt} \simeq  -\frac{1}{1+z} \frac{\vartriangle z}{\vartriangle t}
\end{equation}
therefore,  $H(z)$ can be found directly from Eq.~\eqref{eq:28} once $\frac{dz}{dt}$ is known ~\cite{sei}. Using the present value of the scale factor  normalised to unity , \emph{i.e.}, $a = a_0 =1$ we get a relation of the Hubble parameter with the redshift parameter $z$. If $\rho_0$ be the density at present epoch  then the well known density parameter was written previously as   $\Omega_m = \frac{c}{\frac{BK}{M} + c}$ ~\cite{seth}. Now using Eq.~\eqref{eq:12}, we can express the three space matter density as
\begin{equation}\label{eq:29}
\rho = \rho_0 \left \{1-\Omega_m + \Omega_m \left(1+z \right)^{\frac{M(1+ \alpha)}{2 -dm}} \right \}^{\frac{1}{1+\alpha}}
\end{equation}
\begin{figure}[ht]
\begin{center}
  \includegraphics[width=8 cm]{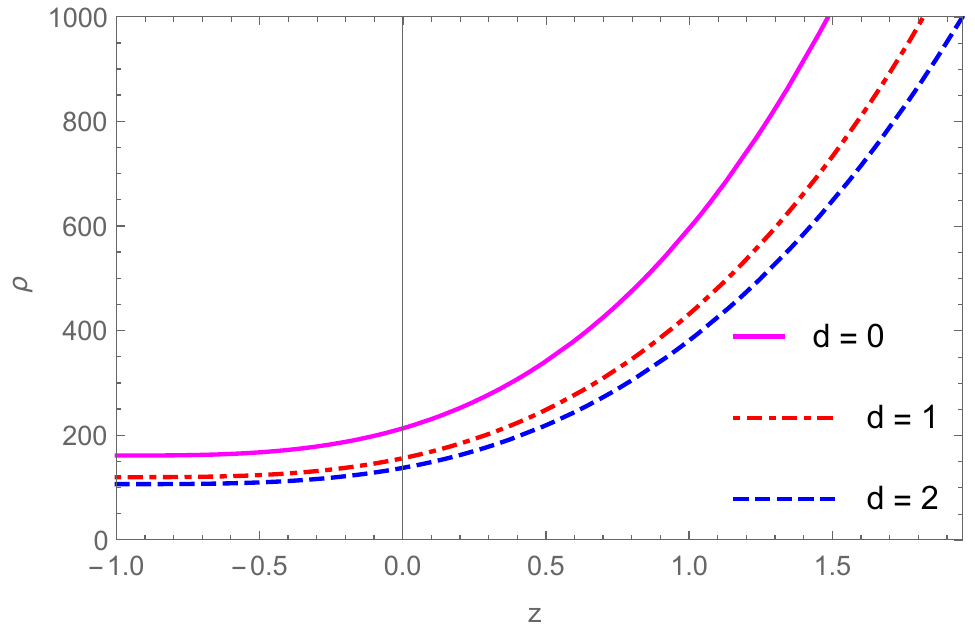}
  \caption{
  \small\emph{The variation of $\rho$ with Redshift  $z$  }\label{fig:rzh71}
    }
\end{center}
\end{figure}
The Fig.-\ref{fig:rzh71} shows that the density of universe at the present epoch in the framework of a higher dimensional is less compared to a $4D$ universe.
Now the Hubble parameter
\begin{equation}\label{eq:30}
 H(z) = H_0 \left \{1-\Omega_m + \Omega_m \left(1+z \right)^{\frac{M(1+\alpha)}{2 -dm}} \right \}^{\frac{1}{2(1+\alpha)}}
\end{equation}
where $H_0 = \left(\frac{2 \rho_0}{k} \right)^{\frac{1}{2}}$ is the present value of the Hubble parameter.
The Eq.~\eqref{eq:30} shows the evolution of Hubble parameter $H (z)$ as a function of redshift parameter $z$.  We draw a best fit curve  of redshift against Hubble parameter in the $1 \sigma$ confidence region  using Hubble $57$ data points in Fig.-\ref{fig:BFG}.

The apparently small uncertainty of the measurement naturally increases its weightage in
estimating $\chi^2$ statistics. We define here the   $\chi^2$ as
\begin{equation}\label{eq:31}
\chi_{H}^2 = \sum_{i=1}^{30} \frac{[H^{obs}(z_i) - H^{th} (z_i, H_0, \theta]^2}{\sigma^2_H(z_i)}
\end{equation}
where $H^{obs}$ is the observed Hubble parameter at $z_i$ and $H^{th}$ is the corresponding
 theoretical Hubble parameter  given by Eq.~\eqref{eq:30}. Also, $\sigma_H(z_i)$ denotes the uncertainty
 for the $i_{th}$ data point in the sample and $\theta$ is the model parameter.
  In this work, we have used the  observational $H(z)$ dataset consisting of $57$
   data points in the redshift range, $0.07 \leq z \leq  2.36$~\cite{dp3}, larger than the redshift
   range that is covered by the type Ia supernova. It should be noted that the confidence levels
 $1 \sigma(68.3\%)$,  $2 \sigma(95.4\%)$  and $3 \sigma(99.7\%)$ are taken proportional to
$\triangle \chi^2 = 2.3 $, $6.17$ and $11.8$ respectively, where $\triangle \chi^2 = \chi^2(\theta) - \chi^2 (\theta* )$   and  $\chi^2_m$ is the minimum value of $\chi^2$. An important quantity which is used for data fitting process is
\begin{equation}\label{eq:32}
\overline{\chi^2} =  \frac{\chi_m^2}{dof}
\end{equation}
where subscript {\it dof} represents the  degree of freedom, and it is defined as the difference between all observational data points and the number of free parameters. If  $\frac{\chi_m^2}{dof} \leq 1$, we get a good fit and the observed data are consistent with the considered model.
\begin{figure}[ht]
\centering
\subfigure[ $d = 0$  ]{
\includegraphics[width= 3 cm]{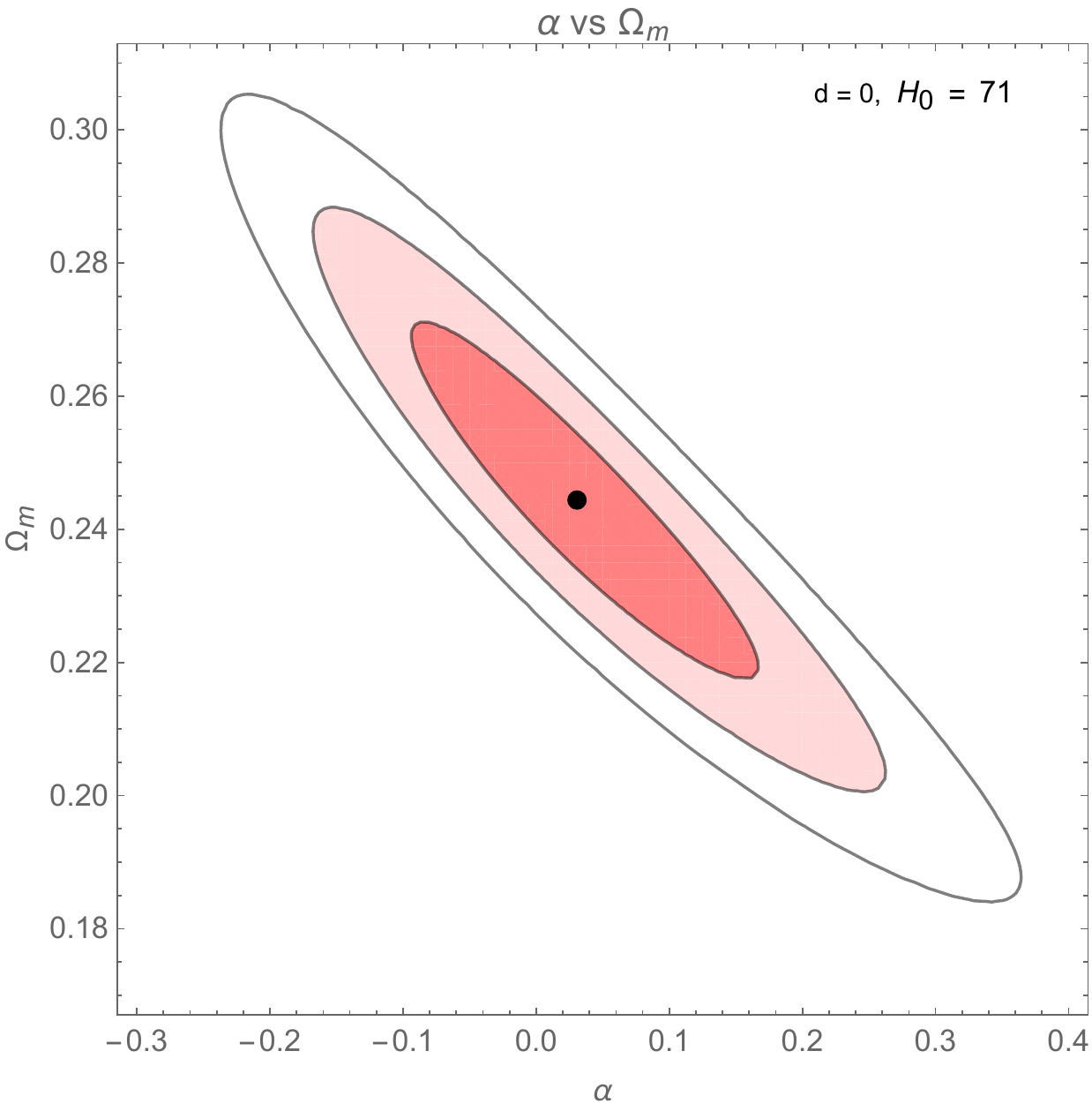}
\label{fig:d071} } ~~~
\subfigure[   $d=1$  ]{
\includegraphics[width= 3 cm]{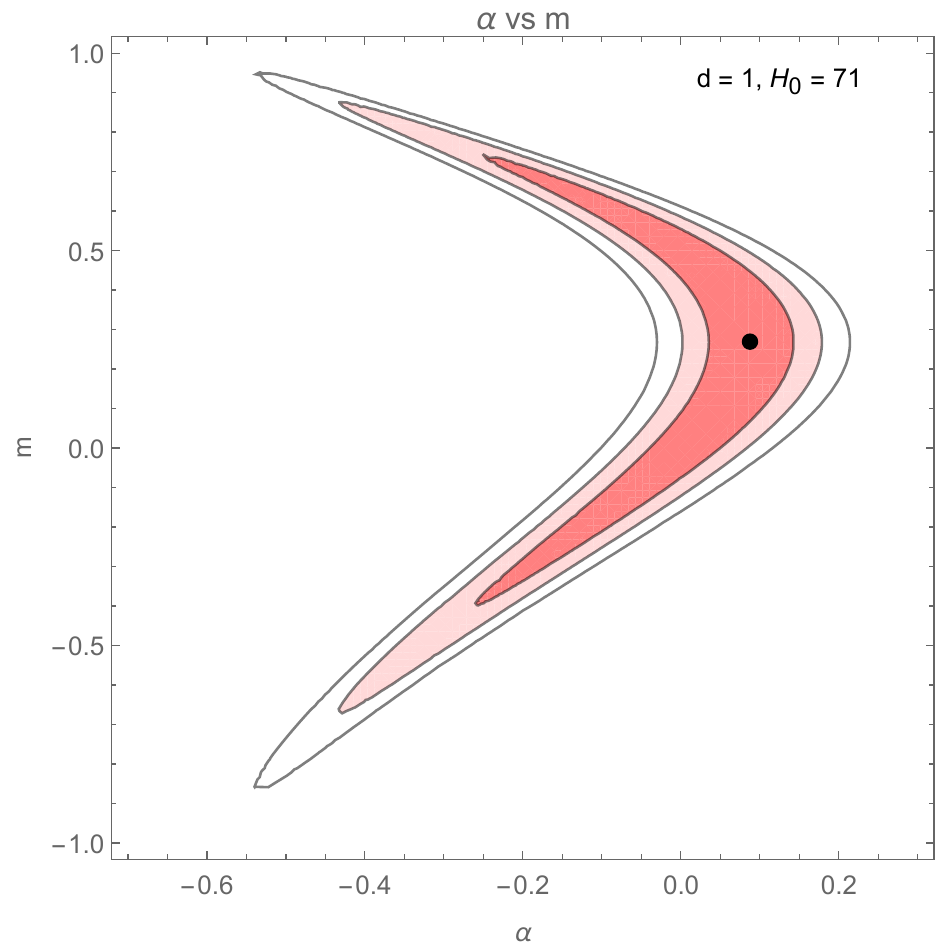}
\label{fig:d171} }
 \subfigure[   $d=2$  ]{
\includegraphics[width= 3 cm]{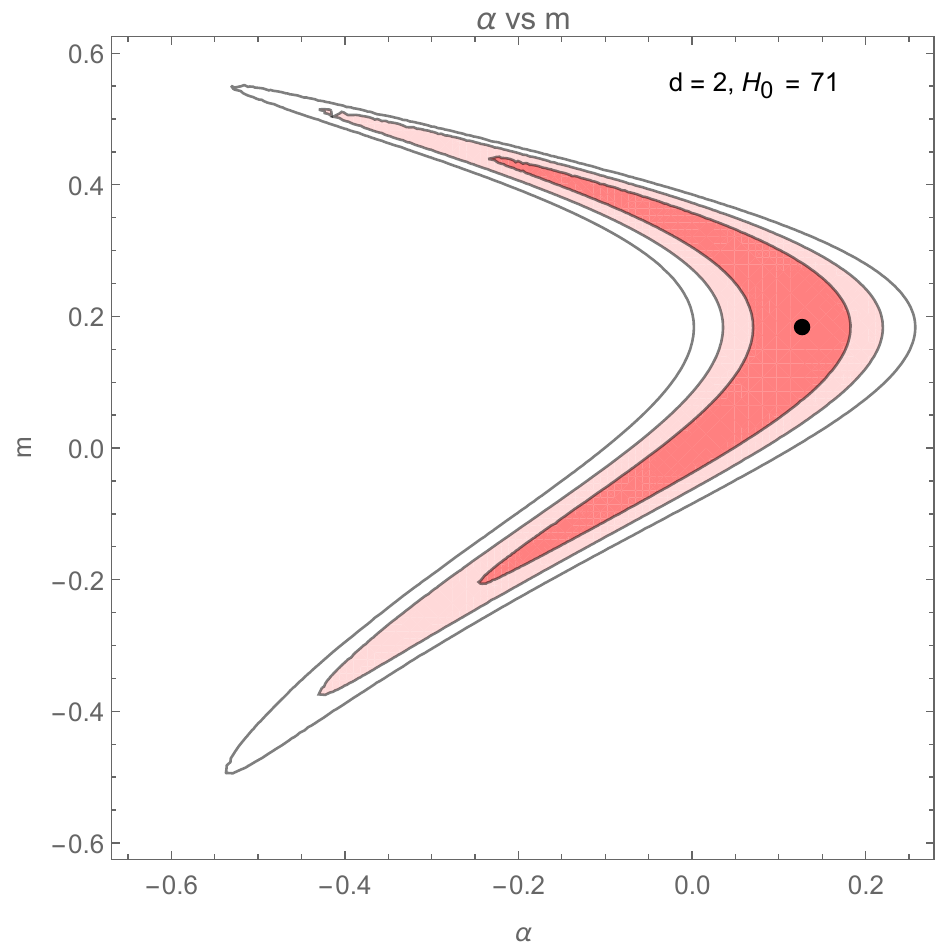}
 \label{fig:d271} }
  ~~~~~~~~~~~\caption[Optional
caption for list of figures]{\emph{ The contour graph of $H(z)$ vs $z$. }}
\end{figure}
We draw the contour graphs using the Hubble-$57$ dataset and obtain the best-fit values of the model parameters at $\chi^2_{M}$ for $H_0 = 71$ km s$^{-1}$ Mpc$^{-1}$~\cite{sah}.
The values of the model parameters $\alpha$, $\Omega_m$, and $m$ corresponding to $d=0,1,$ and $d=2$ are collectively listed in Table-\ref{t1}.
It is worth mentioning that Malekjani \textit{et al}.~\cite{male} obtained a value of $\alpha = 0.033$ in $4$-dimensional spacetime, which is nearly identical to our result.
From Table-\ref{t1} it is shown that the value of $m <1$  for $d = 1$ and it is also less than $0.55$ for $d = 2$ which satisfy our previous conditions $m < \frac{3d - \sqrt{3d(d + 2)}}{d (d - 1)}$ for $d \neq 1 $ and $m < 1$ for $d =1 $. Again, $\chi_{m}^2$ is a statistical measure used to determine how well the model fits the data. Lower values of $\chi_{m}^2$
indicate a better fit. Here, the $\chi_{m}^2$  values are shown for each model, and they indicate how closely the respective model (with the best-fit values of $\alpha$ and $m$ )  matches the Hubble dataset.  The model with $d = 2$ offers a marginally better fit than the $d=0~ \&~ d = 1$ model, as indicated by the lower $\chi^2_m$ value.

\begin{table}[h!]
\centering
\caption{\small Best-fit values of the parameters with $H_0 = 71 ~(\mathrm{km\,s^{-1}\,Mpc^{-1}})$ using Hubble-$57$ data (1$\sigma$ confidence region).}

{\fontsize{08}{10}\selectfont
\begin{tabular}{lccccc}
\hline
\textbf{Parameter} & Dimension ($d$) & $\chi^2_m$ & $\Omega_m$ & $\alpha$ & $m$ \\
\hline
\textbf{Best-fit values} & ~~0~~ & ~~44.77~~ &~~ 0.2443~~ & ~~0.03~~ &~~ - ~~\\
\textbf{Range (1$\sigma$ region)} & ~~ -  ~~ & ~~ - ~~& ~~0.2180, 0.2710~~ &~~ $-0.094, 0.167 ~~$ &~~ - ~~ \\
\textbf{Best-fit values} &~~1 ~~ &~~  44.51 ~~&~~   -  ~~ & ~~0.089 & $0.268$~~ \\
\textbf{Range (1$\sigma$ region)} &~~ - ~~ &~~ - ~~ &~~  - ~~&~~ $-0.2605, 0.1431$ ~~& ~~ $-0.3954, 0.7284$ ~~ \\
\textbf{Best-fit values} &~~2 ~~&~~ 44.36 ~~&~~ - ~~&~~ 0.126 & $0.183$ ~~\\
\textbf{Range (1$\sigma$ region)} &~~ - ~~&~~  -   ~~ &~~  -  ~~&~~  $-0.2405, 0.1844$ ~~&~~ $-0.2056, 0.4465$ ~~ \\
\hline
\end{tabular}
}
\label{t1}
\end{table}

The contours show  that the range of $m$ can be both positive and negative in the $1 \sigma$ region. To get a dimensional reduction of extra dimensions, we consider the positive value of $m$ only.  It is noted that the values of $m$ decrease as the number of dimensions increase, in order to satisfy the condition $2 < dm$. On the other hand, we consider only positive values of $\alpha$ since we are studying with the generalized Chaplygin gas model. Furthermore, the results indicate that $\alpha < 1$ ($\alpha \neq 1$), which contradicts the pure Chaplygin gas model ($\alpha = 1$). However, this finding aligns with our analytical analysis, ensuring that the speed of sound remains less than the speed of light. Additionally, smaller values of $\alpha$ are favored in the late universe.\\

In a higher dimensional model with extra $d$-dimensions it is noted that comparing the data obtained by the differential age method (DA), the model with Chaplygin gas favours a $6D$ universe for a given value of $m$. It is to be mentioned here that from Eq.~\eqref{eq:5} in the framework of  dimensional reduction the 4-dimensional scale factor increases. But we can not explain the impact of compactification of extra dimensions  on the present acceleration of the universe.\\
As pointed earlier  the key Eq.~\eqref{eq:14} is not amenable to  an explicit solution  which is a function of time in known simple form. In this case the variation of cosmological variables like sale factor, flip time, dependence on extra dimensions  etc. can not be explicitly obtained. To avoid such a difficulty of obtaining solution in known form to  determine
the flip time and other physical features of cosmology we adopt here an \emph{an approximate solution method}  in the next section.

\vspace{0.2 cm}

\section{ An approximate solution ansatz :}\label{sec:approx}
 \vspace{0.2 cm}
In the late evolution, the universe is large enough, and the second term on the right-hand side (RHS) of Eq.~\eqref{eq:14} becomes almost negligible compared to the first term. We know that the generalized Chaplygin gas equation of state explains the transition from the dust-dominated era to the present accelerating universe. Since the $4D$ scale factor is sufficiently large, it is reasonable to consider only the first-order approximation in the binomial expansion of the RHS of Eq.~\eqref{eq:14}. We then obtain an \textit{exact solution} for this first-order approximation in the Eq.~\eqref{eq:14}.  Now from Eq.~\eqref{eq:14} we determine the late stage of evolution of the universe neglecting the higher order terms which is given by

\begin{equation}\label{eq:33}
\frac{k}{2}H^2 = \frac{k}{2}\frac{\dot{a}^2}{a^2} =  \left(\frac{Bk}{M} \right)^{\frac{1}{1+\alpha}} + \frac{1}{1+\alpha} \left(\frac{M}{BK}\right)^{\frac{\alpha}{1+\alpha}} \frac{c}{a^{\frac{M(1+\alpha)}{2-dm}}}
\end{equation}
The Eq.~\eqref{eq:33} yields  as first integral an expression of the scale factor as
\begin{equation}\label{eq:34}
a(t) = a_{0} \sinh^{n} \omega t
 \end{equation}
where, $a_{0} =  \left \{ \frac{cM}{ Bk(1 + \alpha)} \right \}^{\frac{2-dm}{M(1 + \alpha)}}$ ; $n = \frac{2(2-dm)}{M(1 + \alpha)}$ and $\omega = \left( \frac{BM^{1+2\alpha}}{k^{\alpha}} \right)^{\frac{1}{2(1 + \alpha)}} \frac{1 + \alpha}{\sqrt{2}(2-dm)}$.
From Eq.~\eqref{eq:34}, it follows that the evolution of both the $3D$ scale factor $a(t)$ and the extra dimension $b(t)$ crucially depend on the values of the exponent $n$.

\begin{figure}[ht]
\begin{center}
  \includegraphics[width=8 cm]{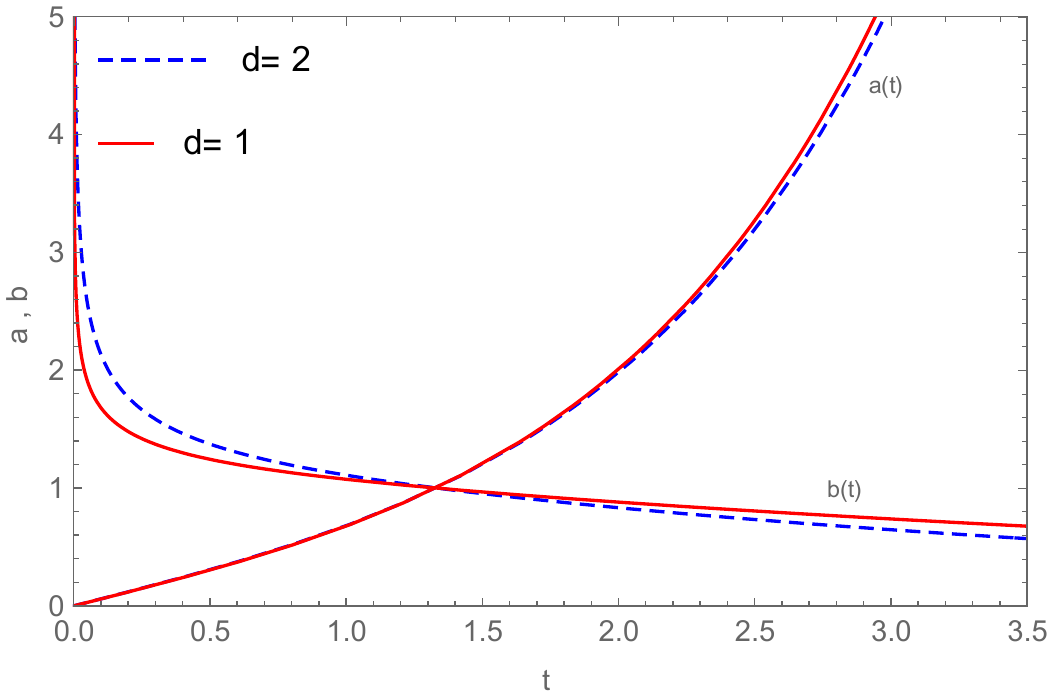}
  \caption{
  \small\emph{The graph involves the variation of $a(t)$ and $b(t)$ with $t$ using equation \eqref{eq:34} }\label{fig:abta}
    }
\end{center}
\end{figure}

In Fig.-\ref{fig:abta},  it is evident that the evolution of the scale factor $a (t)$  and the reduction of extra dimensional scale factor $b(t)$ with time $t$ is determined by different values of $d$. This shows that the desirable feature of dimensional reduction of extra scale factor is possible. It is also seen that the rate of growth of $4D$ scale factor depends on the number of dimensions and it is slower for $ d = 2$.  Again the reduction rate of extra dimensional scale factor is faster for $d = 1$. So  it is physically realistic to consider the presence of extra dimension which may correspond to late acceleration.

Now using Eqs.~\eqref{eq:7}, \eqref{eq:8} and \eqref{eq:34} we can write the expression of $p$ and $\rho$ as follows.

\begin{equation}\label{eq:35}
\rho =  \left(\frac{Bk}{M} \right)^{\frac{1}{1+\alpha}} \coth^2\omega t = \left(\frac{BK}{M}\right)^{\frac{1}{1+\alpha}} \left \{  (1+z)^{\frac{2}{n}} +1 \right\}
 \end{equation}
and

\begin{dmath}\label{eq:36}
p = -  \left(\frac{BM^{\alpha}}{k^{\alpha}} \right)^{\frac{1}{1+\alpha}} \left(1+\alpha - \alpha \coth^2\omega t \right)\\
= -\left(\frac{BM^{\alpha}}{k^{\alpha}} \right)^{\frac{1}{1+\alpha}} \left \{1  -\alpha   (1+z)^{\frac{2}{n}}   \right\}
\end{dmath}
The effective equation of state is given by

\begin{equation}\label{eq:37}
w_e = \frac{p}{\rho} = - \frac{M}{k} \left\{(1+\alpha) \tanh^2 \omega t -\alpha \right\}
= -\frac{M}{k} \frac{1 - \alpha (1+z)^{\frac{2}{n}}}{1 + (1+z)^{\frac{2}{n}}}
\end{equation}
The Eq.~\eqref{eq:37} presents the following results:
\begin{itemize}
    \item[(i)] For a dust-dominated universe, i.e., $ w_e = 0$ , we obtain $z = \frac{1}{\alpha^{\frac{n}{2}}} - 1$. This expression requires  $\alpha < 1 $ for positive $z $. Additionally, for a pure Chaplygin gas where  $\alpha = 1$, we find  $z = 0$, which suggests that, if we observe the universe from a dust-dominated time scale, it appears as if the universe was filled with a pure Chaplygin gas.
    \item[(ii)] At the present epoch (\textit{i.e.}, at $ z = 0$), we find $ w_e = - \frac{M}{k}(1 - \alpha) $, which implies an accelerating universe and it is $-0.9949$ (approx.) for $d =0$. Once again, for a pure Chaplygin gas ($ \alpha = 1 $), we recover a dust-dominated universe ( $w_e = 0 $) at the present epoch, in agreement with the earlier context.
    \item[(iii)] On the other hand, in the late universe, we find $ w_e = - \frac{M}{k} = -1 - \frac{2dm(m+1)}{k} $, which matches the eq.~\eqref{eq:24}. This yields $ w_e < -1 $ for $ d > 0 $, indicating a phantom-type universe. However, for a 4D universe (\textit{i.e.}, $ d = 0 $), we obtain $ w_e = -1$, representing a $\Lambda$CDM model as discussed in Sec.~\ref{sec:eos}.
\end{itemize}

\begin{figure}[ht]
\begin{center}
  \includegraphics[width=8 cm]{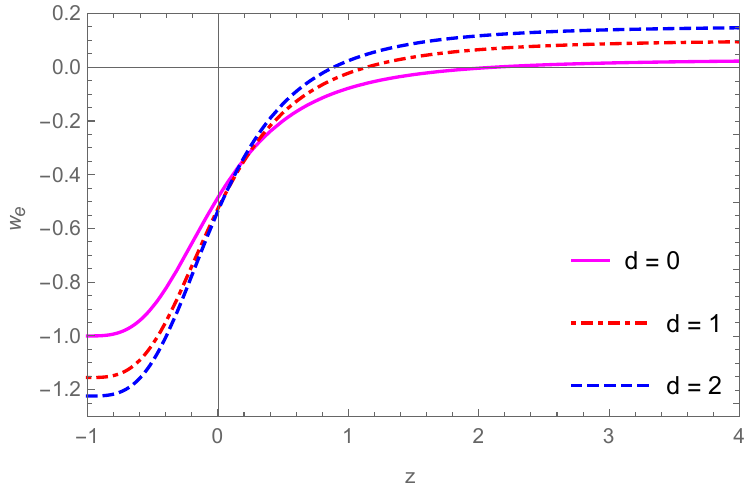}
  \caption{
  \small\emph{The variation of $w_e$ with $z$ using equation \eqref{eq:37} }\label{fig:waz71}
    }
\end{center}
\end{figure}

Now using  Eqs.~\eqref{eq:15} and \eqref{eq:34}  we get the deceleration parameter as

\begin{equation}\label{eq:38}
q = \frac{1-n \cosh^2 \omega t}{n \cosh^2 \omega t}
= \frac{1}{n\left\{1 +(1+z)^{-\frac{2}{n}}\right\}}-1
 \end{equation}
The Eq.~\eqref{eq:38} provides the exponent $n$, which determines the evolution of the deceleration parameter $q$. A numerical analysis using Eq.~\eqref{eq:38} reveals the following:

\begin{itemize}
    \item[(i)] In the early universe, \textit{i.e.} at high $z$, $ q = \frac{1}{n} - 1 $, this corresponds to a dust-dominated universe. For example, in our $4D$ world, $ n = \frac{2}{3(1+\alpha)} $, and $ q \approx 0.5 $ (from Table-\ref{t2}), which is in very good agreement with well-known present $4D$ results. We obtained the same result in Sec.~\ref{sec:q};

    \item[(ii)] At the present epoch (\textit{i.e.}, $ z = 0 $), Eq.~\eqref{eq:27} reduces to $ q = \frac{1}{2n} - 1 $. In the 4D world, $ q \approx  - 0.247 $ (from Table-\ref{t2}), which corresponds to an accelerating universe. It is notable that $q$  approaches $-1$  from the past towards the present epoch, consistent with a transition to acceleration over cosmic time;

    \item[(iii)] In the late universe, we find $ q = -1 $, which represents a pure $\Lambda$CDM model with no dimensional dependence observed here.  These same results for the present and future epochs were also shown in Sec.~\ref{sec:q}.
\end{itemize}
\begin{figure}[ht]
\begin{center}
  \includegraphics[width=8 cm]{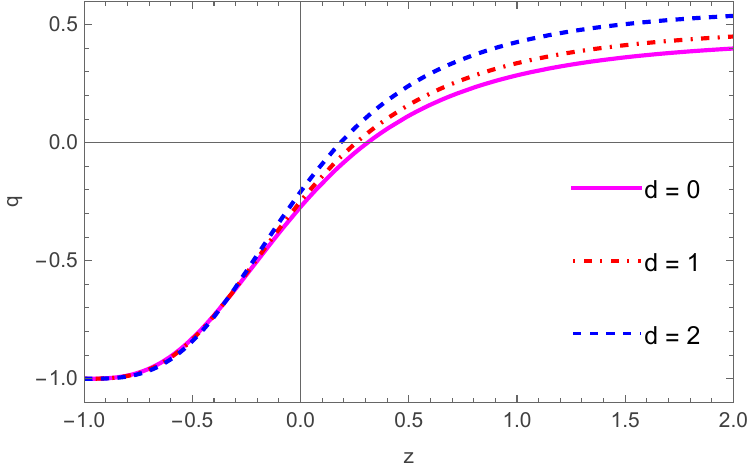}
  \caption{
  \small\emph{The variation of $q$ with $z$ using equation \eqref{eq:38} }\label{fig:qaz71}
    }
\end{center}
\end{figure}
Again, Fig.-\ref{fig:qaz71}  shows the variation of $q$ with $z$. The deceleration parameter is higher for higher dimensions. It is also seen that the flip occurs later  for more dimensions like Fig.-\ref{fig:qzh71}.\\
Now the flip time ($t_f$) is given by,
\begin{equation}\label{eq:39}
t_f = \frac{1}{\omega} \cosh^{-1} \left(\sqrt{\frac{1}{n}} \right)
\end{equation}
 which is the time at which the deceleration parameter $q$ changes sign, marking the transition from a decelerating to an accelerating phase in the universe's expansion.
 \begin{figure}[ht]
\begin{center}
  \includegraphics[width=8 cm]{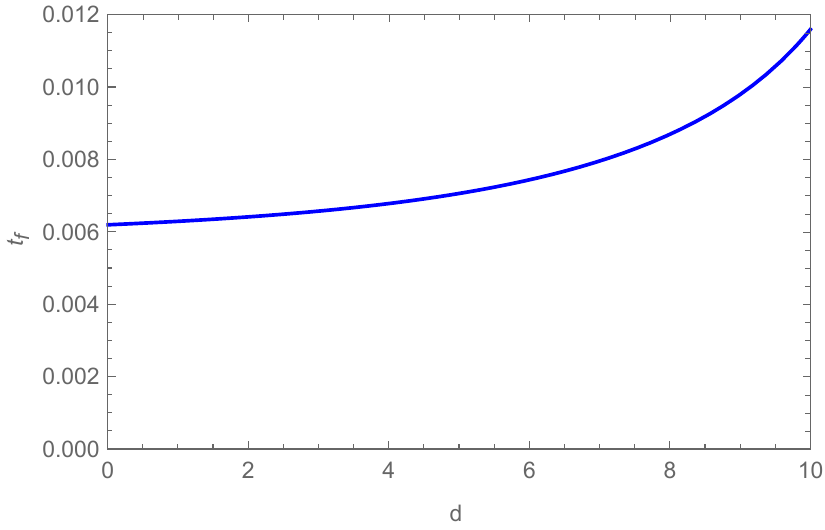}
  \caption{
  \small\emph{The variation of $t_f$ with $d$ is shown} \label{fig:tfd}
    }
\end{center}
\end{figure}
From Fig.-\ref{fig:tfd}, it is observed that the flip occurs at a later time as the number of dimensions increases. This indicates that higher dimensions tend to delay the transition from deceleration to acceleration. This behavior may be attributed to the modified gravitational dynamics in higher-dimensional cosmological models, where the additional spatial dimensions influence the effective expansion rate of the Universe. Such an effect is consistent with theoretical expectations in higher-dimensional cosmology.

The transition time $t_f$, which marks the change from a decelerating to an accelerating phase of cosmic expansion, is governed by the parameter $n$, which depends inversely on $(1+\alpha)$. Consequently, the Chaplygin gas parameter $\alpha$ plays a crucial role in determining the timing of the accelerated phase of the Universe. For $0<\alpha<1$, the parameter $n$ takes relatively larger values, which in turn reduces the flip time $t_f$. This implies that the transition to accelerated expansion occurs at an earlier cosmological epoch. In contrast, for $\alpha=1$, corresponding to the original Chaplygin gas model, the value of $n$ becomes smaller, leading to an increase in $t_f$. As a result, the acceleration phase is realized comparatively later in cosmic history.

This behavior indicates that smaller values of $\alpha$ shift the transition to acceleration toward earlier times, which is in better agreement with observational constraints on the transition redshift $z_f$ given in Eq.~\eqref{eq:17}. The present analysis is consistent with our earlier results, where it was shown that a decrease in $\alpha$ leads to a higher transition redshift $z_f$ and a smaller flip time $t_f$. Both effects imply that the Universe enters the accelerated expansion phase earlier for lower $\alpha$, while still preserving the characteristic feature that acceleration remains a late-time phenomenon.

Now, using Eq.~\eqref{eq:38}, the redshift parameter at the flip is given by
\begin{equation}\label{eq:40}
z_{fa} = \left(\frac{n}{1-n} \right)^{\frac{n}{2}} -1 .
\end{equation}
This expression requires $n<1$ for $z_{fa}$ to be real, which is consistent with observational constraints. To achieve acceleration at the present epoch, we must have $z_{fa}>0$, which further leads to the condition $ 4 > \frac{M(1+\alpha)}{2-dm} > 2(1+\alpha)$. In the four-dimensional case (\textit{i.e.}, $d=0$), Eq.~\eqref{eq:40} reduces to the simple requirement $\alpha<\tfrac{1}{3}$. For acceleration at the present epoch, this condition must be satisfied (in our case $\alpha=0.0051$), which lies well within the observationally allowed range and is consistent with current cosmological data. In contrast, $\alpha=1$ is not admissible in this scenario since it violates the constraint $\alpha<\tfrac{1}{3}$.

Now the expression of jerk parameter is given by
\begin{dmath}\label{eq:41}
j = \left[\frac{1}{n \left \{1 + (1+z)^{-\frac{2}{n}} \right \} }- 1 \right]\left[\frac{2}{n \left \{1 + (1+z)^{-\frac{2}{n}} \right \} }- 1 \right] \\
 + \left[\frac{2(1+z)^{-\frac{2}{n}}}{n^2 \left \{1 + (1+z)^{-\frac{2}{n}} \right \}^2 }  \right]
\end{dmath}
\par
Since the Eq.~\eqref{eq:41} is so involved  we shall discuss  the nature of jerk parameter  by graphical presentation.
\begin{figure}[ht]
\begin{center}
  \includegraphics[width=8 cm]{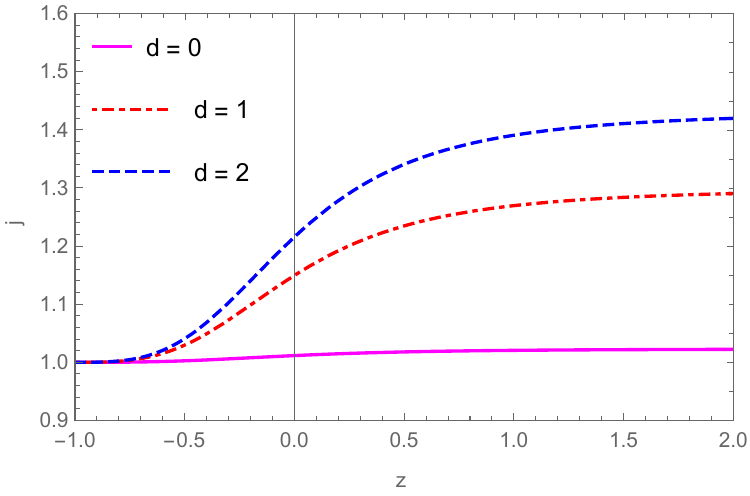}
  \caption{
  \small\emph{The variation of $j$ with $z$ using equation \eqref{eq:41} }\label{fig:jaz71}
    }
\end{center}
\end{figure}
The graph in Fig.-\ref{fig:jaz71} shows three distinct lines, each representing a different dimensions $d = 0$,  $d= 1$ and $d=2$. The lines are converging at $j = 1$ which corresponds to the
$\Lambda$CDM  model at a future cosmic time. This convergence is in good agreement with current observational data about our universe. It suggests that at late universe the evolution
favours a $4D$ world.
Now the expression of Hubble parameter in terms of $\Omega_m$ is given by

\begin{equation}\label{eq:42}
H = H_0 \Omega_m^{\frac{1}{2(1+\alpha)}}\left \{\left(\frac{1 - \Omega_m}{\Omega_m}\right)^{\frac{1}{1+\alpha}}  + \frac{1}{1 + \alpha}\left(\frac{\Omega_m}{1 - \Omega_m}\right)^{\frac{\alpha}{1+\alpha}}
(1+z)^{\frac{M(1+\alpha)}{2 - dm}}\right\}^{\frac{1}{2}}
\end{equation}

\begin{figure}[ht]
\centering
\subfigure[ $d = 0$  ]{
\includegraphics[width= 3 cm]{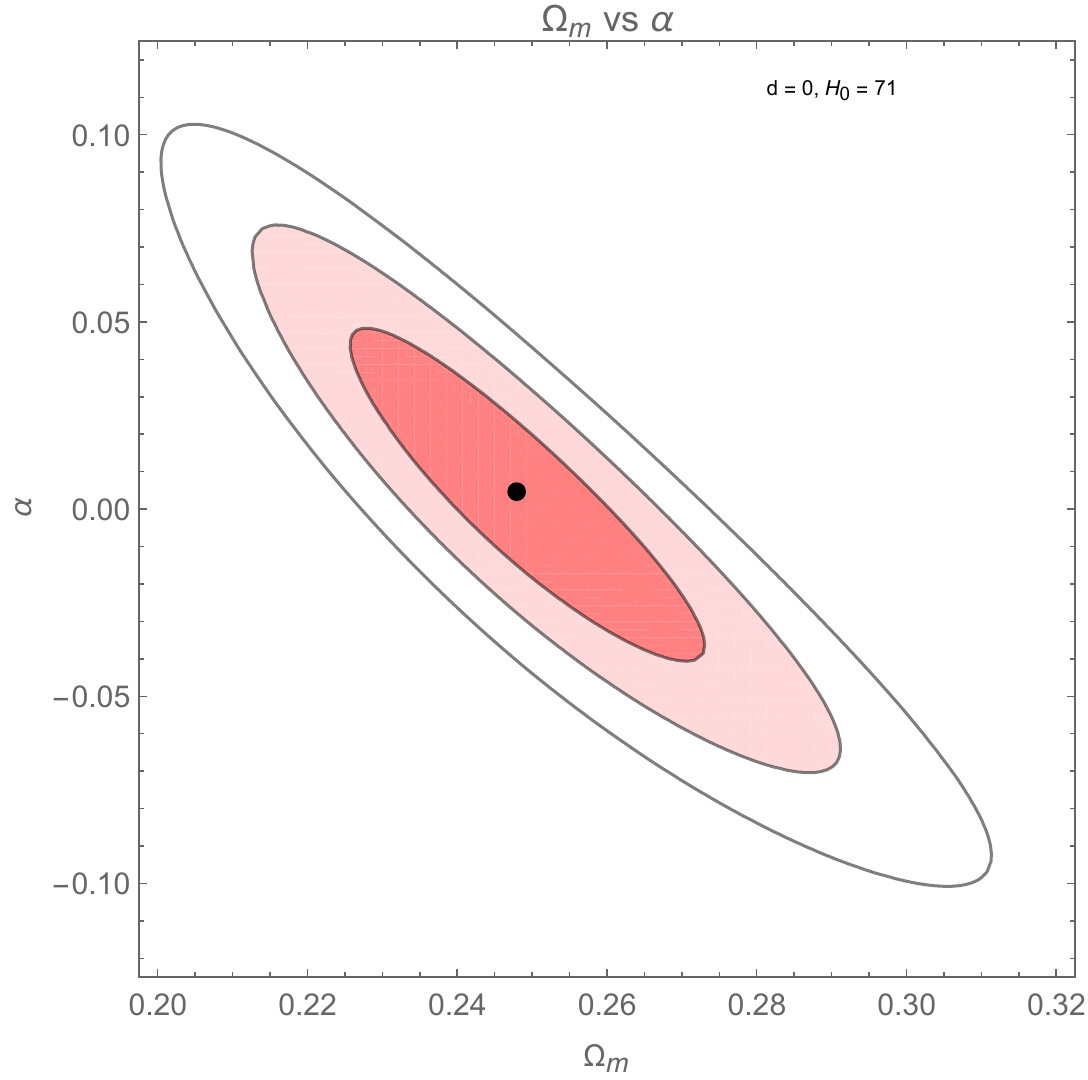}
\label{fig:d0h71} } ~~~
\subfigure[   $d=1$  ]{
\includegraphics[width= 3 cm]{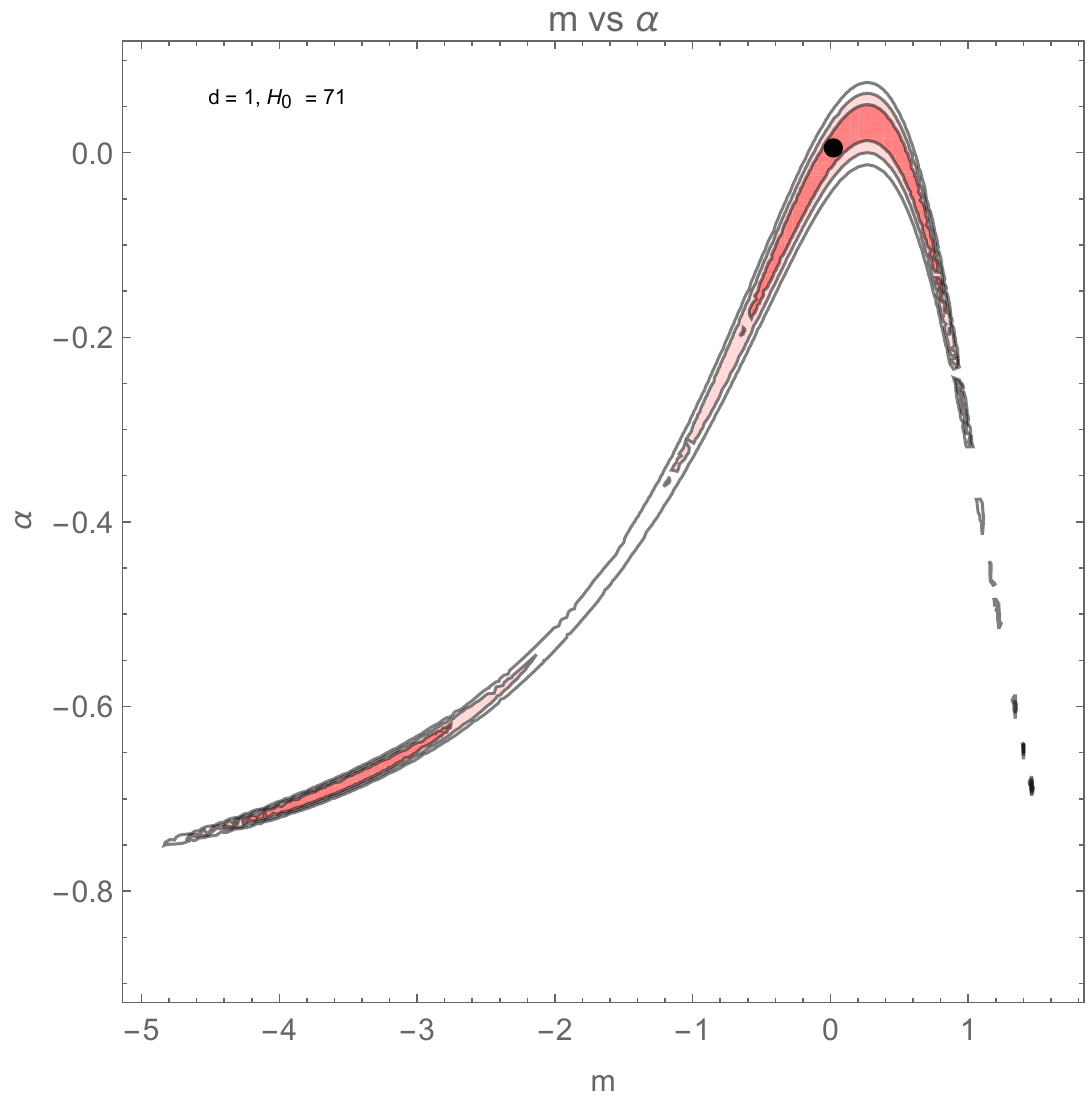}
\label{fig:d1h71} }
 \subfigure[   $d=2$  ]{
\includegraphics[width= 3 cm]{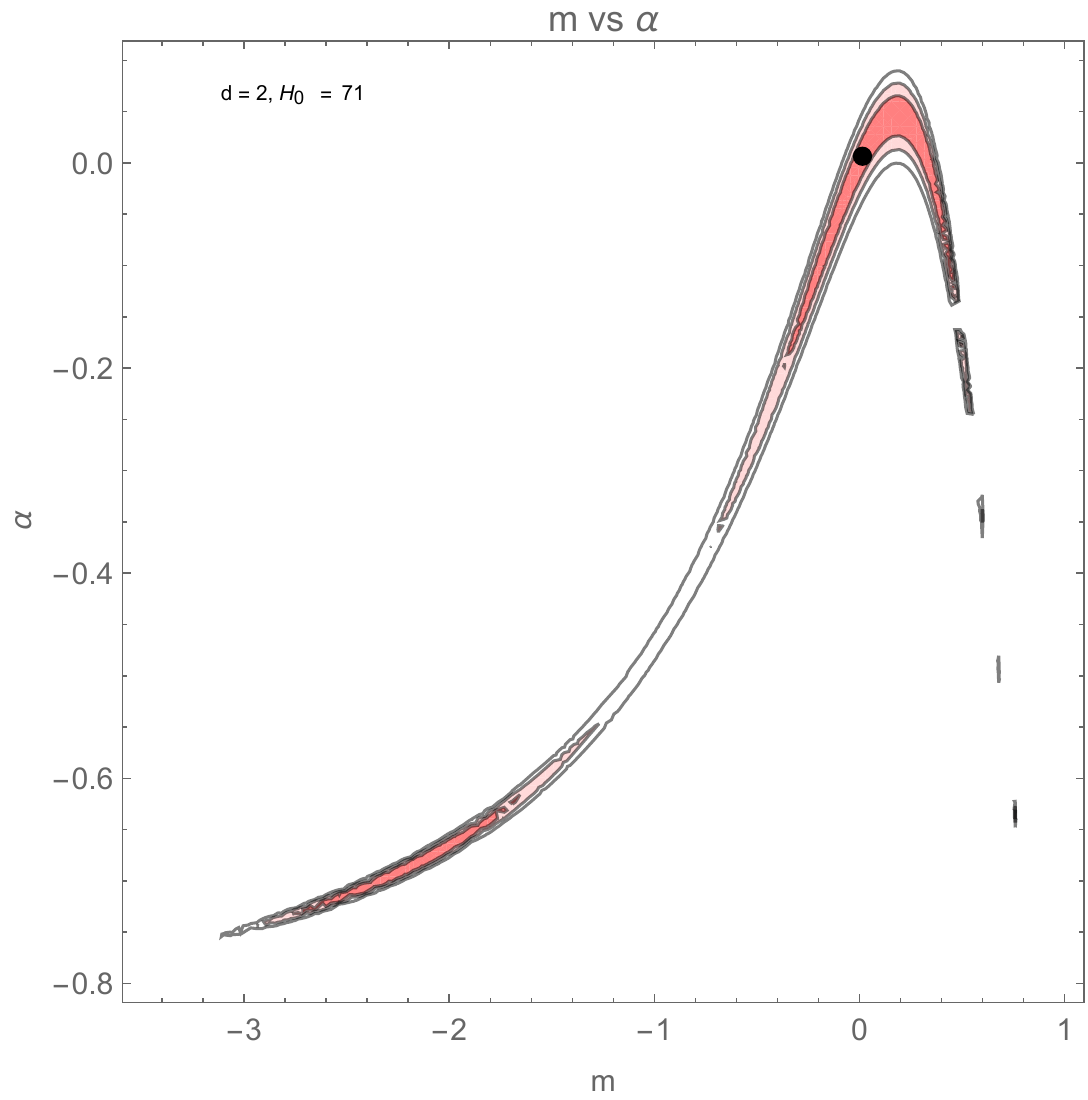}
 \label{fig:d2h71} }
  ~~~~~~~~~~~\caption[Optional
caption for list of figures]{\emph{ The contour graph of $H(z)$ vs $z$. }}
\end{figure}
\begin{table}[h!]
\centering
\caption{\small Best-fit values of the parameters with $H_0 = 71 ~(\mathrm{km\,s^{-1}\,Mpc^{-1}})$ using Hubble-$57$ data (1$\sigma$ confidence region).}

{\fontsize{08}{10}\selectfont
\begin{tabular}{lccccc}
\hline
\textbf{Parameter} & Dimension ($d$) & $\chi^2_m$ & $\Omega_m$ & $\alpha$ & $m$ \\
\hline
\textbf{Best-fit values} & ~~0~~ & ~~44.87~~ &~~ 0.248~~ & ~~0.0045~~ &~~ - ~~\\
\textbf{Range (1$\sigma$ region)} & ~~ -  ~~ & ~~ - ~~& ~~0.2262, 0.2719~~ &~~ $-0.0410, 0.0470 ~~$ &~~ - ~~ \\
\textbf{Best-fit values} &~~1 ~~ &~~  44.89 ~~&~~   -  ~~ & ~~0.0055 & $0.0215$~~ \\
\textbf{Range (1$\sigma$ region)} &~~ - ~~ &~~ - ~~ &~~  - ~~&~~ $-0.1799, 0.0525$ ~~& ~~ $-0.5989, 0.3083$ ~~ \\
\textbf{Best-fit values} &~~2 ~~&~~ 44.89 ~~&~~ - ~~&~~ 0.0055 & $0.0106$ ~~\\
\textbf{Range (1$\sigma$ region)} &~~ - ~~&~~  -   ~~ &~~  -  ~~&~~  $-0.1865, 0.1871$ ~~&~~ $-0.3714, 0.0674$ ~~ \\
\hline
\end{tabular}
}
\label{t2}
\end{table}

Here, we draw the contour graphs using the Hubble-$57$ dataset and obtain the best-fit values of the model parameters at $\chi^2_{m}$ for $H_0 = 71$ km s$^{-1}$ Mpc$^{-1}$~\cite{sah}. The values of $\alpha$, $\Omega_m$, and $m$ for the cases $d=0,1,$ and $d=2$ are listed in Table-\ref{t2}.
To constrain the parameters, let us consider $B_s = \frac{B(1+ \alpha)}{\frac{BK}{M}(1+\alpha) + c} $, we get the expression of Hubble parameter in terms of $B_s$ as

\begin{equation}\label{eq:43}
H = H_0 \left \{ \frac{B_s k}{M} +\left(1- \frac{B_s k}{M} \right)\left(1+z \right)^{\frac{M(1+\alpha)}{2 - dm}}
\right\}^{\frac{1}{2}}
\end{equation}

\begin{figure}[ht]
\centering
\subfigure[ $d = 0$  ]{
\includegraphics[width= 3 cm]{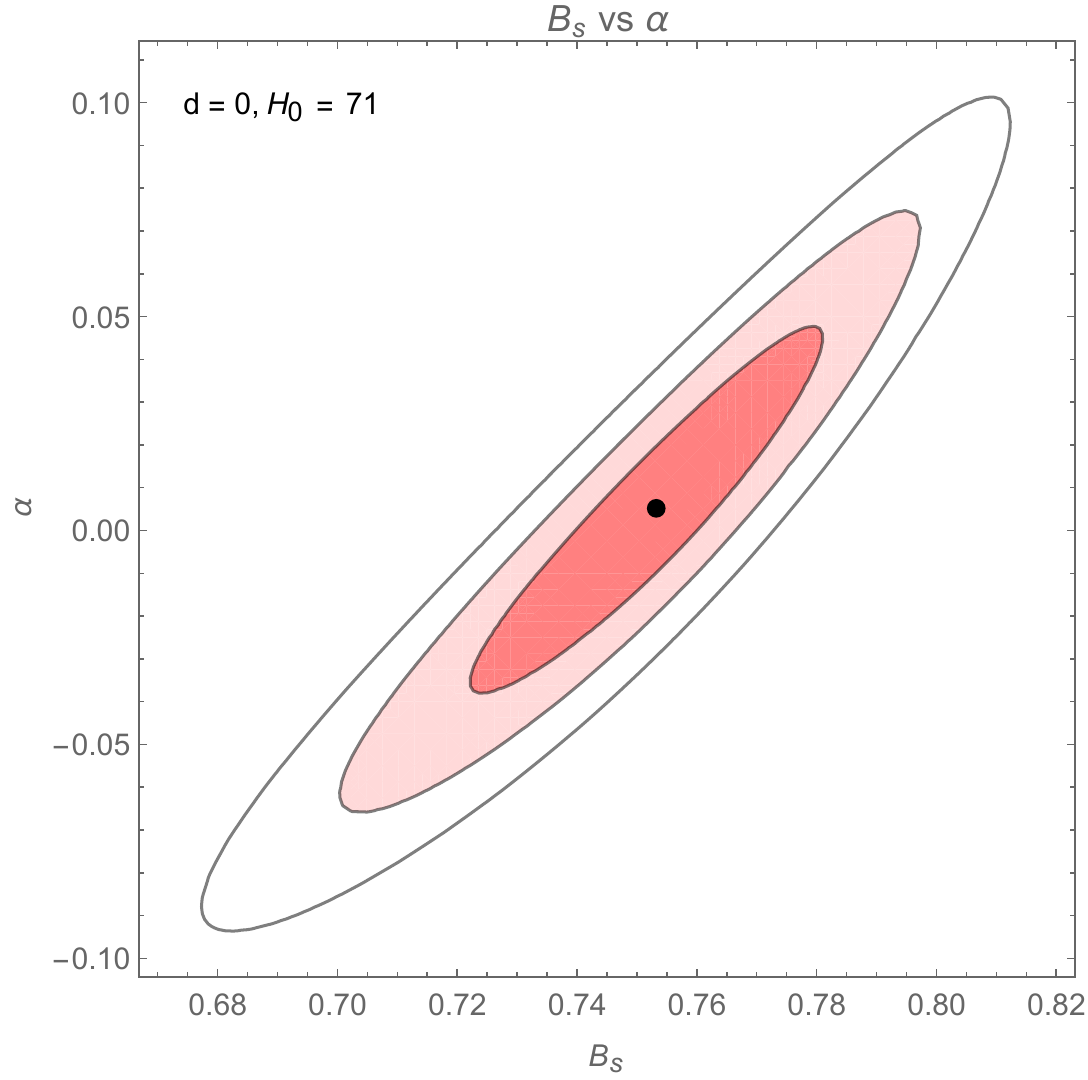}
\label{fig:da071} } ~~~
\subfigure[   $d=1$  ]{
\includegraphics[width= 3 cm]{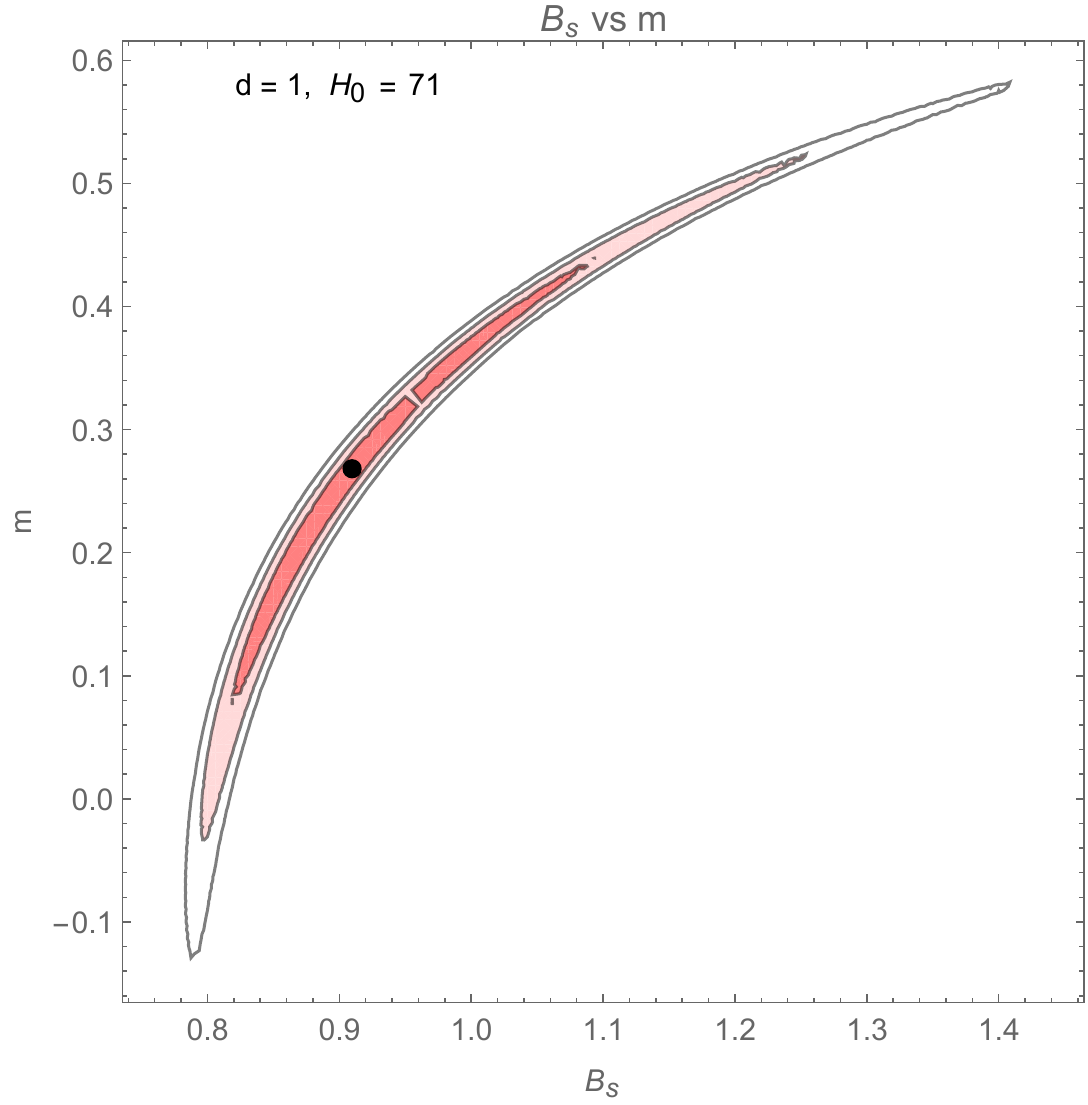}
\label{fig:da171} }
 \subfigure[   $d=2$  ]{
\includegraphics[width= 3 cm]{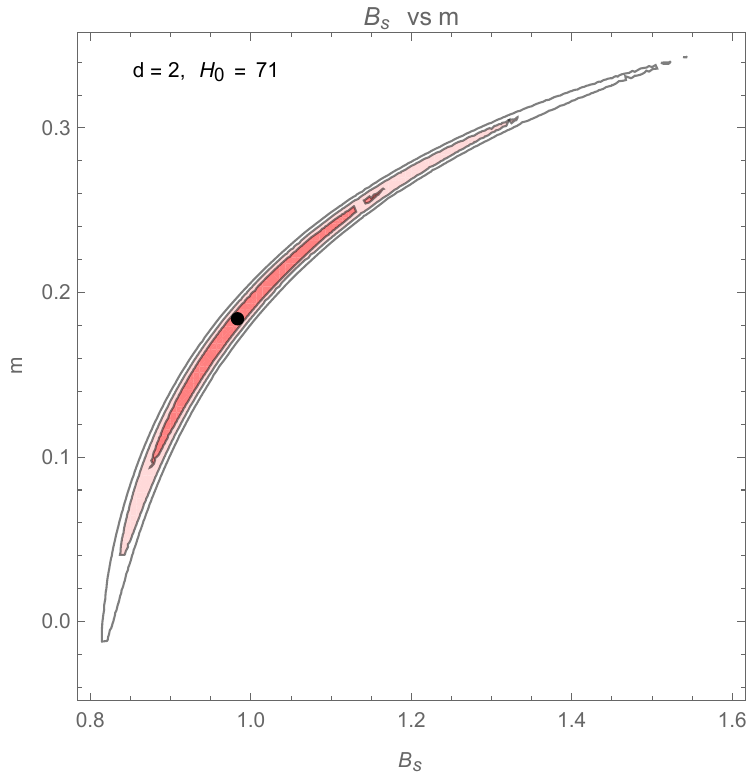}
 \label{fig:da271} }
  ~~~~~~~~~~~\caption[Optional
caption for list of figures]{\emph{ The contour graph of $H(z)$ vs $z$. }}
\end{figure}
The values of $B_s$ and $\alpha$ at $\chi^2_m$ for $d=0$, together with the corresponding values of $\alpha$ and $m$ at $\chi^2_m$ for $d=1$ and $d=2$, are extracted from the contour graphs and summarized in Table-\ref{t3}. We find that $\alpha<1$, in contrast to the pure Chaplygin gas model where $\alpha=1$.

\begin{table}[h!]
\centering
\caption{\small Best-fit values of the parameters with $H_0 = 71 ~(\mathrm{km\,s^{-1}\,Mpc^{-1}})$ using Hubble-$57$ data (1$\sigma$ confidence region).}

{\fontsize{08}{10}\selectfont
\begin{tabular}{lccccc}
\hline
\textbf{Parameter} & Dimension ($d$) & $\chi^2_m$ & $B_s$ & $\alpha$ & $m$ \\
\hline
\textbf{Best-fit values} & ~~0~~ & ~~44.86~~ &~~ 0.7532~~ & ~~0.0051~~ &~~ - ~~\\
\textbf{Range (1$\sigma$ region)} & ~~ -  ~~ & ~~ - ~~& ~~0.7226, 0.7807~~ &~~ $-0.0372, 0.0475 ~~$ &~~ - ~~ \\
\textbf{Best-fit values} &~~1 ~~ &~~  44.05 ~~&~~   0.9100 ~~ & ~~  -  ~~ &~~$0.268$~~ \\
\textbf{Range (1$\sigma$ region)} &~~ - ~~ &~~ - ~~ &~~  0.8212, 1.087 ~~&~~ -  ~~& ~~ $0.0820, 0.4324$ ~~ \\
\textbf{Best-fit values} &~~2 ~~&~~ 53.63 ~~&~~ 0.9820 ~~&~~  - ~~ &~~ $0.183$ ~~\\
\textbf{Range (1$\sigma$ region)} &~~ - ~~&~~  -   ~~ &~~  0.8768, 1.129 ~~&~~  -  ~~&~~ $0.0955, 0.2499$ ~~ \\
\hline
\end{tabular}
}
\label{t3}
\end{table}

Here, we determine the best-fit values of $B_s$ and $m$ by minimizing $\chi^2$, as shown in the previous tables. Notably, the $d = 0$ model fits the data slightly better than the $d = 1$ and $d = 2$ models, which disagrees with the earlier model discussed in Section 3 that represents a 6-dimensional universe. This discrepancy may be attributed to the neglect of higher-order terms in Eq.~\eqref{eq:14}.
From Table-\ref{t2}, it is observed that for $\chi^2_{m}$, we obtain positive values of $B_s$ and $m$ for each value of $d$. The value of $B_s$ is higher for $d = 2$ compared to $d = 1$ and $d = 0$. If we refer to Table-\ref{t3}, we see that the range of $B_s$ in the $1\sigma$ region is positive for $d = 0$, $d = 1$, and $d = 2$. On the other hand, positive values of $m$ are shown here, which favour the dimensional reduction of the extra space.  It is worth noting that the values for $d = 1$ and $d = 2$ are identical for both approaches, and the values of $m$ decrease as the number of dimensions increases, in order to satisfy the condition $2 < dm$. In Eq.~\eqref{eq:43}, we have used the corresponding value of $\alpha$ to draw the contour. The previous tables indicate that $\alpha$ is significantly less than unity, consistent with our earlier approach discussed in Sec.~\ref{sec:obs}.
To ensure that the speed of light is greater than the speed of sound, a lower value of $\alpha$ is expected at the late universe, which is consistent with our observational results. Furthermore, from Table-\ref{t2}, it is also seen that the value of $m$ satisfies the previous conditions related to $m$.

The Generalized Chaplygin Gas (GCG) model has been extensively studied in 4-dimensional spacetime by several authors, who have constrained the model parameters using various observational data. For instance, Malekjani \textit{et al.}\cite{male} obtained a value of $B_s = 0.76$, which closely aligns with our findings for $d = 0$. Similarly, Bertolami \textit{et al.}\cite{bert} determined that $B_s$ ranges from $0.62$ to $0.82$, and $\alpha$ ranges from $0.052$ to $1.056$; P. Wu and H. Yu~\cite{wu} found that $0.67 \leq B_s \leq 0.83$ and $-0.21 \leq \alpha \leq 0.42$. Additionally, P. Thakur~\cite{thak} obtained $B_s = 0.772$ and $\alpha = 0.023$. The values of $B_s$ and $\alpha$ found by these authors in 4-dimensional spacetime are comparable to those obtained through our approximate solution approach for $d = 0$.

\section { Summary :}\label{sec:sum}
In this paper, we present a higher-dimensional cosmological model that uses a generalized Chaplygin gas to explain the recent acceleration of the universe. This work generalizes our previous study~\cite{dpej}, where the pure Chaplygin gas in higher-dimensional spacetime was considered. It also generalizes our recent article~\cite{dp3}, where the generalized Chaplygin gas (GCG) was discussed in four-dimensional spacetime. We obtain new results and highlight the merits  and demerits of the approaches presented in this article. The Eq.~\eqref{eq:14} is the key equation of this work; however, due to its highly nonlinear nature, it cannot be solved exactly and has only been studied under extremal conditions. Additionally, important cosmological parameters—such as the evolution of the three-dimensional scale factor, the dimensional reduction of extra spaces, and the flip time— cannot be explicitly derived from Eq.~\eqref{eq:14}.

To address this shortcoming, we propose an approximation method that neglects the higher-order terms in the binomial expansion of the right-hand side of Eq.~\eqref{eq:14}. This approximation is justified because, in a dust-dominated universe, the scale factor is expected to be sufficiently large, making it reasonable to consider only the first term of the binomial expansion, as demonstrated in Eq.~\eqref{eq:33}.

We analyze the deceleration parameter, effective equation of state (EoS), and the jerk parameter for both the models, applying parameter constraints using Hubble-$57$ data. Key findings are as follows:

\begin{enumerate}[(i)]
    \item The solutions obtained here are general in nature, as they reduce to all the well-known results of 4D Chaplygin gas-driven cosmology when $d = 0$~\cite{dp3}. Furthermore, if we set $\alpha = 1$, all the solutions reduce to those in our previous work~\cite{dpej}. Upon comparing our results with current observations, we find that they favour $\alpha < 1$, which does not align with the pure Chaplygin gas model, but it is consistent with our analysis that the speed of sound is lower than the speed of light in order to maintain causality. It is also found that the value of $\alpha$ becomes smaller during late acceleration.

    \item  Positive values of the parameter $m$  allow dimensional reduction, applicable to both the approaches. Hubble-$57$ data indicate positive $m$ values, as shown in Table-\ref{t1} \& Table-\ref{t2} for our model. It is also found that   the values of $m$ decrease as the number of dimensions increases, in order to satisfy the condition $2 < dm$.

    \item The model parameters obtained from the Hubble-$57$ data using the two approaches are compared in Table-\ref{tp}.

\begin{table}[h!]
\centering
\caption{\small Dimension-wise comparison of the best-fit model parameters obtained from the two approaches.}
{\fontsize{8}{10}\selectfont
\begin{tabular}{c c c c c c c c c c}
\hline
\multicolumn{4}{c}{1st approach} & & \multicolumn{5}{c}{2nd approach} \\
\cline{1-4}\cline{6-10}
$d$ & $\alpha$ & $\Omega_m$ & $m$ & & $d$ & $\alpha$ & $\Omega_m$ & $m$ & $B_s$ \\
\cline{1-4}\cline{6-10}
0 & 0.030 & 0.2443 & --    & & 0 & 0.0045 & 0.248 & --    & 0.7532 \\
1 & 0.089 & --     & 0.268 & & 1 & 0.0055 & --    & 0.268 & 0.9100 \\
2 & 0.126 & --     & 0.183 & & 2 & 0.0055 & --    & 0.183 & 0.9820 \\
\hline
\end{tabular}
}
\label{tp}
\end{table}
Table-\ref{tp} shows that, for each fixed value of the extra-dimensional parameter $d$, both approaches yield consistent estimates of the parameter $m$.
The parameter $\Omega_m$ remains comparable across the two methods for the same $d$, whereas the generalized Chaplygin gas parameter $\alpha$ assumes noticeably smaller values in the second approach.
Furthermore, the approximation-based approach enables a more detailed characterization of the Chaplygin gas sector through the parameter $B_s$, leading to a refined description of the late-time cosmic dynamics.

 \item The most remarkable feature of the approximation method lies in the fact that unlike previous cases here we obtain an \textit{explicit} solution for the $3D$ scale factor $a(t)$, as well as the expression for the extra dimensions $b(t)$, which is shown in Fig.-\ref{fig:abta}. It is noted that the 3D scale factor increases more rapidly for lower dimensions, while the reduction of the extra space scale factor is slower for higher dimensions. The presence of extra dimensions may correspond to late acceleration.

 \item  We have studied the deceleration parameter $q$ with respect to the redshift parameter  $z$.
    \begin{itemize}
        \item[(a)] The GCG model depicts the evolution from a dust-dominated to an accelerating universe. For high $z$, it represents a dust-dominated phase, with $q = \frac{1}{2}$ for $d = 0$, matching the well-known $4D$ universe and indicating deceleration. At the present epoch ($ z = 0 $),  $q \approx -0.625$  for $d = 0$ suggests an accelerating universe. At the late stage of evolution, $ q = -1 $ corresponds to a pure $ \Lambda $CDM model which is independent of extra dimensions. This is due to the dimensional reduction of extra spaces, which have no effects in the late universe, showing very good agreement with our observable universe
        \item[(b)] In the approximation method, the universe also appears dust-dominated at high $ z $. At $ z = 0 $, $ q \approx -0.247 $ for $ d = 0 $ signifies acceleration, and in later epochs, $ q = -1 $ leads to the $\Lambda$CDM model.
    \end{itemize}
    Both approaches yield more or less similar results. Notably, the deceleration parameter increases with higher dimensions, and the flip to accelerated expansion occurs later in models with higher dimensions. This indicates that the presence of extra spatial dimensions delays the onset of present cosmic acceleration, indicating a retarding influence from these additional dimensions on the acceleration of the universe.

  \item The flip time ($t_f$) has been derived using our approximation method in Eq.~\eqref{eq:39}. From Fig.-\ref{fig:tfd}, it is evident that the transition to acceleration occurs at progressively later times as the number of extra dimensions increases. This suggests that higher-dimensional effects retard the transition to cosmic acceleration, likely due to modified gravitational dynamics, in agreement with theoretical expectations that dimensionality can influence cosmic evolution. The transition time $t_f$ is controlled by the parameter $n$, which depends inversely on $(1+\alpha)$. Consequently, the generalized Chaplygin gas parameter $\alpha$ plays a key role in determining when acceleration begins. For $0<\alpha<1$, larger values of $n$ lead to a smaller $t_f$, implying an earlier transition to accelerated expansion. In contrast, for $\alpha=1$ (the original Chaplygin gas case), smaller $n$ results in a larger $t_f$, delaying the acceleration phase. This behavior is consistent with our earlier findings that decreasing $\alpha$ increases the transition redshift $z_f$ (Eq.~\eqref{eq:17}) while reducing $t_f$. Thus, smaller $\alpha$ shifts the acceleration to an earlier—but still late-time—epoch, in better agreement with observational constraints.
\item
\begin{itemize}
        \item[(a)] Eq.~\eqref{eq:17} defines the redshift parameter at the flip time $z_{f}$ which represents the point where the deceleration parameter changes sign. For the universe to be accelerating at the present epoch  (\textit{i.e.}, at $z = 0$), this leads to the condition $z_f >0$, which implies the constraint $\Omega_m < \frac{ 2(2-dm)}{M} < 1$.  In the special case where $d = 0$, this  condition simplifies  to  $\Omega_m < \frac{2}{3}$  which matches current observational constraints on $\Omega_m$, indicating it’s less than about $0.3$ (in our case $\Omega_m = 0.2443$). This alignment with observations supports the idea of an accelerating universe.

           The transition from deceleration to acceleration depends on both the number of extra dimensions $d$ and the Chaplygin gas parameter $\alpha$. For $0<m<1$, the transition redshift $z_f$ decreases with increasing $d$, indicating that higher dimensions delay cosmic acceleration, whereas smaller $\alpha$ shifts the transition to an earlier epoch.

        From Table-\ref{t_zf}, it is evident that $z_f$ is sensitive to both $d$ and $\alpha$. In the four-dimensional case ($d=0$), smaller $\alpha$ yields a significantly larger $z_f$ compared to the original Chaplygin gas model, implying an earlier but still late-time flip to acceleration. As $d$ increases, $z_f$ decreases systematically, confirming that higher-dimensional effects retard the acceleration phase. Overall, the generalized Chaplygin gas model with small $\alpha$ provides a physically viable and observationally consistent description of the recent accelerated expansion of the Universe.
\item[(b)] For the approximation method presented here we also determine the redshift parameter $z_{fa}$ at flip time in Eq.~\eqref{eq:40}, which gives $n < 1$ for real $z_{fa}$, a condition that aligns with observational findings. To achieve acceleration at the present epoch, we require $z_{fa}> 0$, which further imposes the constraint $4 > \frac{M(1+\alpha)}{2-dm} > 2(1+\alpha)$. In the 4-dimensional case (\textit{i.e.}, $d = 0$), Eq.~\eqref{eq:40} simplifies to $\alpha < \frac{1}{3}$. For acceleration at the present epoch, we find that $\alpha < \frac{1}{3}$ (in our case, $\alpha = 0.0051$), a condition that is consistent with values obtained from observational data as well as our theoretical analysis.  In contrast, $\alpha=1$ is not admissible in this scenario since it violates the constraint $\alpha<\frac{1}{3}$.

 \end{itemize}
     \item
    \begin{itemize}
        \item[(a)] Eq.~\eqref{eq:22}  provides the effective EoS in the GCG model. Initially, $ w_{\text{eff}} = 0 $ indicates a dust-dominated universe. At present epoch, $ w_{\text{eff}} \approx -0.75 $ for a $4D$ case implies acceleration. In the later epoch, for $ d = 0 $, $w_{\text{eff}} = -1 $ denotes the $\Lambda$CDM model, while for $ d \neq 0 $, $ w_{\text{eff}} < -1 $ suggests a phantom-type universe, this appears  due to the  effect of extra dimensions. Our present universe with quintessence like Dark Energy(DE) transits to a universe with phantom type DE in future.  In this respect our work recovers the effective equation of state (for large scale factor) for an article of Guo \textit{et al}~\cite{guo} where a very generalised Chaplygin type of gas is taken.
        \item[(b)] In the approximation method, Eq.~\eqref{eq:37}  describes the effective EoS ($w_e $). This expression requires $ \alpha < 1 $  for positive $z$ in the dust-dominated phase. Interestingly, for $\alpha = 1$ we find $ z = 0$, which suggests that if we observe the universe  from dust dominated time scale, it appears as if the universe was filled with Chaplygin gas. In the late universe, $ w_e < -1 $ for $ d \neq 0 $ suggests a phantom phase, while $ w_e = -1 $ for $d = 0 $ gives the $\Lambda$CDM model, as discussed in section 3.2. At the present epoch ($ z = 0 $), $ w_e = -\frac{M}{k}(1 - \alpha) $ implies acceleration and it is $-0.9949$ for $d =0$. For $\alpha = 1 $, $ w_e = 0 $, represents  a dust-dominated universe.
    \end{itemize}
\item A comparative analysis of the deceleration parameter $q$ and the effective equation of state  in four-dimensional spacetime ($d=0$) obtained from both approaches is presented below.
\begin{table}[h!]
\centering
\caption{\small $q$ and effective EoS for $d = 0$.}
{\fontsize{08}{10}\selectfont
\centering \begin{tabular}{c c c c c}
\hline & \multicolumn{2}{c}{1st approach} & \multicolumn{2}{c}{2nd approach} \\
\hline $z$ & $q$ & $w_{\text{eff}}$ & $q$ & $w_e$ \\
\hline
Early epoch & $0.5$ & $0$ & $0.5$ & $0$ \\
Present epoch & $-0.625$ & $-0.75$ & $-0.247$ & $-0.9949$ \\
Late epoch & $-1$ & $-1$ & $-1$ & $-1$ \\
\hline \end{tabular}
}
\label{tqw}
\end{table}
Table-\ref{tqw} shows that, at early times, both approaches consistently describe a dust-dominated Universe with $q=\frac{1}{2}$ and vanishing equation-of-state parameters, namely $w_{\text{eff}}=0$ in the first approach and $w_e=0$ in the second approach. At the present epoch, both models predict accelerated expansion ($q<0$); however, the second approach exhibits a stronger dark-energy behavior, with the equation-of-state parameter $w_e$ lying closer to $-1$ compared to $w_{\text{eff}}$ obtained in the first approach. In the late-time limit, both approaches converge to $q=-1$ along with $w_{\text{eff}}=-1$ and $w_e=-1$, respectively, indicating an exact de Sitter phase consistent with the $\Lambda$CDM model.

    \item  Calculated in Eqs.~\eqref{eq:27} and \eqref{eq:41} and shown in Fig.-\ref{fig:jzh71}   and Fig.-\ref{fig:jaz71}, the jerk parameter $j $ converges to $1$, corresponding to the $\Lambda $CDM model at a future cosmic time, aligning with current observational data.

    \item  The GCG model in higher-dimensional space-time demonstrates that cosmological dynamics at the late stage are dimension-independent due to the dimensional reduction of extra space. Additionally, in higher dimensions, the flip occurs later, meaning extra spatial dimensions slow down the acceleration in both approaches.

    Since lower values of  $\chi_m^2 $ indicate a better fit, the model with $ d = 2$ provides a marginally better fit than $ d = 1$, as shown by the lower $ \chi_m^2$  values. However, in our approximation method,  $\chi_m^2$ is lower for $ d = 0$, which disagrees with the previous approach discussed in Sec.~\ref{sec:obs}. This discrepancy may result from neglecting higher-order terms in Eq.~\eqref{eq:14}.

    Considering the approximation in the second approach, we may conclude that the higher-dimensional GCG model favors a 6-dimensional universe at the early stage, as per Hubble-$57$ data. Nevertheless, the approximation method remains valuable as it elucidates the evolution of the scale factor, the dimensional reduction of extra space, flip time and more.
   Once again, the role of extra dimensions is found to be insignificant in influencing the late-time acceleration of the universe in a higher-dimensional (HD) framework. However, it is observed that in  a higher-dimensional universe transitions from an initial decelerating phase to a late-time accelerating phase take place at a later epoch compared to a 4-dimensional universe.
\end{enumerate}

\textbf{Acknowledgment : }

DP acknowledges financial support of Netaji Nagar Day College for a Minor Research project.


\begin{thebibliography}{40}
\bibitem{pd} T. Padmanabhan  - `Understanding our Universe : Current status and open issues'  and references
therein, gr- qc/0503107
\end{thebibliography}
\end{document}